\documentclass[twocolumn,showpacs,prb]{revtex4}
\usepackage{amsmath}
\usepackage{epsf}

\newcommand{\beq}{\begin{equation}}
\newcommand{\eeq}{\end{equation}}
\newcommand{\bea}{\begin{eqnarray}}
\newcommand{\eea}{\end{eqnarray}}

\begin{document}

\title{Specific heat of a one-dimensional interacting Fermi system: the role of
anomalies}
\author{Andrey V. Chubukov$^1$, Dmitrii L. Maslov$^2$, and Ronojoy Saha$^{3,4}$}

\begin{abstract}
We re-visit the issue of the temperature dependence of the specific heat $%
C(T)$ for interacting fermions in 1D. The charge component $C_c
(T)$ scales linearly with $T$, but the spin component $C_s (T)$
displays a more complex behavior  with $T$ as it depends on the
backscattering amplitude, $g_1$, which scales down under RG
transformation  and eventually behaves as $g_1 (T) \sim 1/\log T$.
We show, however, by direct perturbative calculations  that $C_s
(T)$ is strictly linear in $T$ to order $g^2_1$ as it contains the
renormalized backscattering amplitude not on the scale of $T$, but
at the cutoff scale set by the momentum dependence of the
interaction around $2k_F$. The running amplitude $g_1 (T)$ appears
only at third order and gives rise to an extra $T/\log^3 T$ term
in $C_s (T)$. This agrees with the results obtained by a variety
of bosonization techniques. We also show how to obtain the same
expansion in $g_1$  within the sine-Gordon model.
\end{abstract}

\affiliation{$^1$Department of Physics, University of
Wisconsin-Madison, 1150 Univ. Ave., Madison, WI 53706-1390}

\vspace{0.5cm}
\affiliation{$^2$Department of Physics, University of Florida, P.
O. Box 118440, Gainesville, FL 32611-8440}

\vspace{0.5cm}
\affiliation{$^3$Institute for Physical Science and Technology and
Department of Physics, University of Maryland, College Park, MD
20742}

\vspace{0.5cm}
\affiliation{$^4$Department of Physics and Materials Science
Institute, University of Oregon, Eugene, OR 97403}
\maketitle

\section{Introduction}

The hallmark of a Fermi liquid is the linear dependence of the specific heat
$C(T)$ on temperature. A deviation from linearity at the lowest temperatures
generally implies  a non-Fermi liquid behavior. This generic rule is
satisfied in dimensions $D >1$, e.g., the non-Fermi-liquid behavior  near
quantum critical points is characterized by a divergent effective mass and
sub-linear specific heat.  On the other hand, the behavior of the best
studied non-Fermi liquids -- one-dimensional (1D) systems of fermions  -- is
more subtle. A 1D system of fermions  can be mapped onto a system of 1D
bosons. As long as these bosons are free, i.e., the system is in the
universality class of a Luttinger liquid,  the specific heat is linear in $T$
despite  that  other properties of a system show a manifestly
non-Fermi-liquid behavior. However, backscattering and Umklapp scattering of
original fermions give rise to interactions among bosons.  If these
interactions are marginally irrelevant, $C(T)$  may acquire an additional $%
\log T$ dependence.

In series of recent publications, several groups studied specific heat of
interacting Fermi systems in  dimensions $1<D\leq 3$  [%
\onlinecite{old,all_1,all_3,cm,all_4,cmgg,cmm,chm,ae,chm_cooper}]. These
systems are Fermi liquids, and the leading term is $C(T)=\gamma T$. The
subleading term is, however, non-analytic: it scales as $A_{D} T^{D}$ (with
an extra $\log T$ factor in 3D), and  in $1<D <3$ the prefactor is expressed
exactly via the spin and charge components of the fully renormalized
backscattering amplitude \cite{cmgg,ae,chm_cooper}
\begin{equation}
A_{D}= - a_{D} \left(\frac{m^*}{k_F}\right)^2~\left( f_{c}^{2}(\pi
)+3f_{s}^{2}(\pi )\right)  \label{1}
\end{equation}
where $a_{D}$ is a number [$a_2 = 3\zeta (3)/(2\pi)$], and $f_{s}(\pi )$ and
$f_{c}(\pi )$ are components of the backscattering amplitude $f(\theta = \pi)
$  ($\theta $ is the angle between the incoming momenta).  The spin and
charge contributions to the specific heat can be extracted independently by
measuring the specific heat at zero and a finite magnetic field (a strong
enough magnetic field $\mu _{B}H\gg T$ reduces the spin contribution to 1/3
of its value in zero field).

As $D\rightarrow 1$, $T^{D}$ becomes $T$, and the universal subleading term
in the specific heat becomes comparable to the leading term. In addition,
the spin component of the backscattering amplitude in 1D  flows under a
renormalization group (RG) transformation, and, for a repulsive interaction,
which is the only case studied in this paper, scales as $1/\log T$ in the
limit $T\rightarrow 0$~ [\onlinecite{giamarchi_book}]. The charge component,
$f_{c}(\pi ),$ on the other hand, remains finite. Judging from Eq. (\ref{1}%
), one might then expect that the charge component of the specific heat in
1D scales as $T$, while the spin component, $C_s(T)$, scales as $T/\log ^{2}T
$.

This simple argument is, however, inconsistent with recent result  obtained
by Aleiner and Efetov (AE) [\onlinecite{ae}]  for the model of weakly
interacting electrons. They  developed a powerful ``multidimensional
bosonization'' method in which fermions are integrated out and the action is
expressed solely in terms of interacting, low-energy bosonic modes. In 1D,
AE showed that $C_s (T)$ behaves as $T/\log^3(T)$ for $T\to 0$ (in
disagreement with the RG argument), and that the logarithmic flow of $f_s$
shows up in $C(T)$ only at fourth order in the interaction.  Similar results
have been previously obtained for the Kondo model \cite{kondo} and XXZ spin $%
1/2$ chain \cite{lukyanov}, which  are believed to be in the same
universality class as 1D fermions with repulsive interaction. [Earlier
perturbative studies of $C(T)$ in 1D yielded different results:  in Ref.~[%
\onlinecite{halperin}], $C(T)$ was argued to be linear in $T$ to all orders
in the interaction, whereas Ref.~[\onlinecite{nersesyan}] found that $C_s(T)$
scales as $T/\log ^{2}T$, both results are in disagreement with  the result
by AE.]

The functional form of $C_s (T)$ is not a purely academic issue.
In a strong enough magnetic field $\mu_B H \gg T$ the $\log T$
term is replaced by the $\log H$ one.  Measuring the field
dependence of $C(T, H)$, one can explicitly determine the
functional form of $C(T, H=0)$. We note in passing that the issue
of universal temperature corrections to thermodynamic quantities
is not restricted to the specific heat.  Number of researchers
studied the universal temperature and wavevector dependence of the
spin susceptibility~ \cite{chi}. Another example of a universal,
non-analytic
behavior  is the $T \sqrt{H}$ behavior of the specific heat  of a 2D $d-$%
wave superconductor in a magnetic field~\cite{volovik}.

Absence of the logarithmic renormalization of $C_{s}(T)$ below
fourth order of perturbation is a rather non-trivial result in
view of Eq. (\ref{1}), but even more so because backscattering in
1D contributes to the specific heat already at first order in the
interaction (see Sec.\ref{sec:first}). Moreover, both first and
second-order  contributions to $C_s (T)$  can be straightforwardly
obtained in a computational scheme in which they  appear as
contributions from low energies, of order $T$.  In the RG spirit,
one might expect these terms to contain the running backscattering
amplitude at a scale of order $T$. However, in 1D, the existence
of a particular computational scheme, in which the answer comes
from low energies, does not actually guarantee that the
corresponding coupling is a running one, as  1D systems with a
linear spectrum are well-known to exhibit anomalies, similar to
Schwinger terms in current-current commutation relations.

From computational viewpoint, the anomaly-type  contribution to $C(T)$ can
be equally obtained either as a low-energy contribution, or as a
contribution from high energies, of the order of the cutoff.  In the latter
case, the corresponding  coupling is on the scale of the cutoff, rather than
$T$. One then has to explicitly evaluate higher-order terms to verify
whether the  coupling is a bare one or a running one.

This running vs. bare coupling dilemma was discussed actively in
the earlier days of bosonization \cite{twocutoffs,grest,solyom},
and is related to a more general issue of how to treat properly
the high-energy cutoffs in theories with linear
dispersions~\cite{volovik_gut}.

Our interest in  the 1D problem is three-fold. First, we want to understand
which of the backscattering couplings entering $C(T)$ are the running ones
and which are the bare ones. We argue below that anomaly-type terms  should
be treated as high-energy contributions, for which the  couplings are at the
cutoff scale. The running coupling appears in $C(T)$  due to
non-anomalous contributions, which can be uniquely identified as  low-energy
contributions.
Second, we would like to check directly whether the $g$-ology model
is a renormalizable theory or not, i.e., whether the dependence of the
ultraviolet cutoffs can be incorporated into a finite (and small) number of
renormalized vertices.
Third, we want to establish parallels between the direct
perturbative expansion in the backscattering amplitude  in momentum space,
and the real-space calculations within the sine-Gordon model. In particular,
we want to understand how  anomaly-type contributions appear in real-space
calculations. This has not been considered in earlier works~\cite
{ae,lukyanov,cardy,ludwig_cardy} for which the main interest was a search
for a contribution with the running coupling. 

\subsection{Model}

\begin{figure}[tbp]
\begin{center}
\epsfxsize=1.0\columnwidth \epsffile{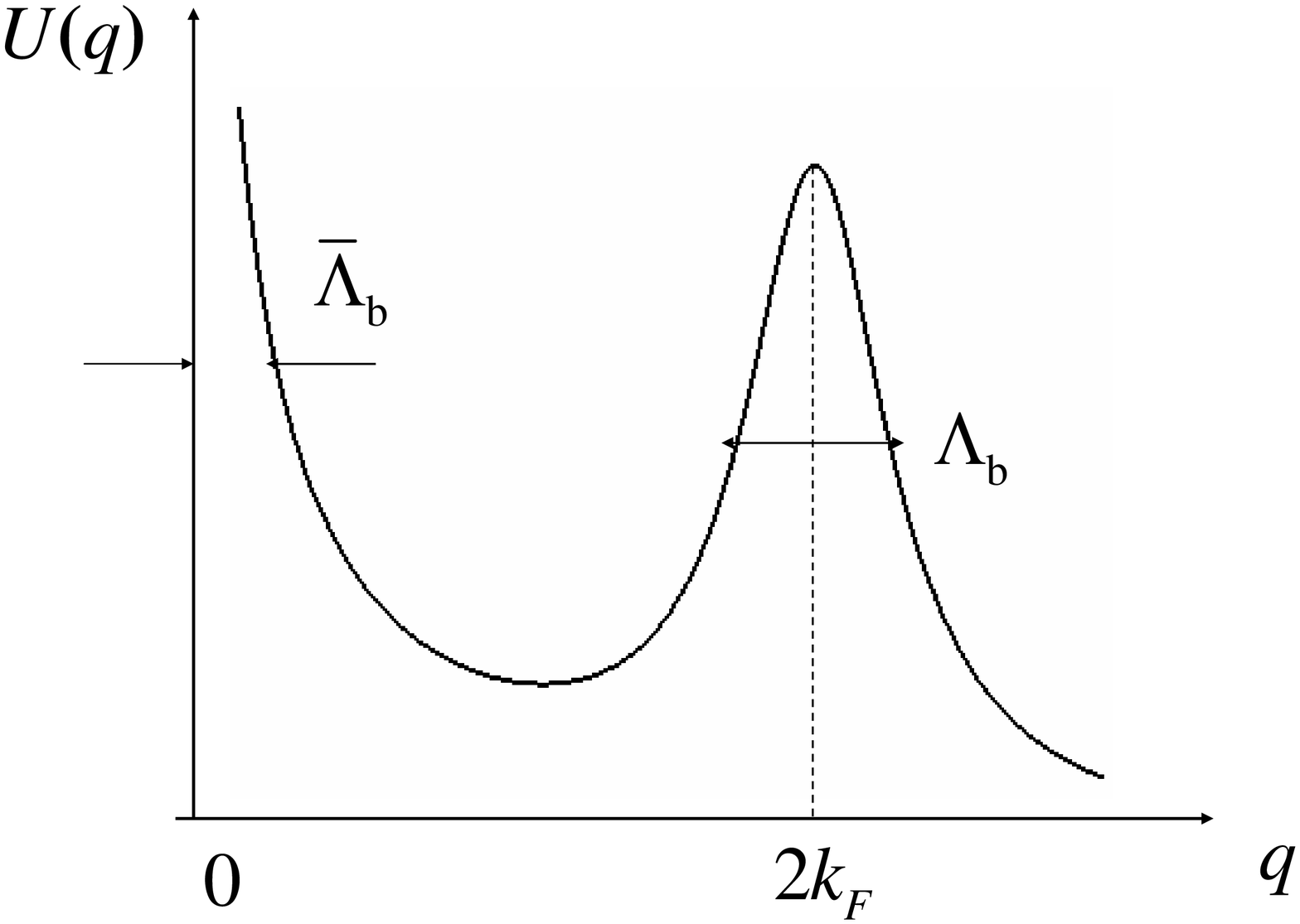}
\end{center}
\caption{A model interaction potential with two cutoffs: ${\bar \Lambda}_{%
\mathrm{b}}$ near $q=0$ and $\Lambda_{\mathrm{b}}$ near $q=2k_F$. }
\label{fig:potential}
\end{figure}
We consider an effective low-energy  model of 1D fermions with a linearized
fermionic dispersion $\epsilon _{k}$ near $\pm k_{F}$, $\epsilon _{k}=\pm
v_{F}(k\mp k_{F})$, and with a short-range four-fermion interaction $U(q)$.
We set a fermionic momentum cutoff at a scale 
$\Lambda_{\mathrm{f}}$ (generically, comparable to the lattice constant)
, and assume that fermions with energies larger than $v_F \Lambda_{\mathrm{b}%
}$ account for the renormalization of the bare interaction into an effective
one, which acts between low-energy fermions, and  depends not only on
transferred momentum, but also on two incoming fermionic momenta. We then
use the $g$-ology notations~\cite{giamarchi_book},  and introduce three
dimensionless vertex functions, $g_{1} $, $g_{2}$ and $g_{4}$, which
describe scattering processes along the Fermi surface with zero incoming and
$2k_{F}$ transferred momenta, zero incoming and zero transferred momenta,
and $2k_{F}$ incoming and zero transferred momenta, respectively. At first
order in the interaction, $g_{1}=U(2k_{F})/(2\pi v_{F})$, $%
g_{2}=g_{4}=U(0)/(2\pi v_{F})$.
The effective low-energy  model only makes sense if the couplings $g_i$
vanish long before the scale of $\Lambda_{\mathrm{f}}$, otherwise the
low-energy and high-energy sectors could not be separated. A  way to enforce
this constraint, which we will adopt, is to assume  that the interactions $%
g_i$ are non-zero only for  transferred momenta (either around zero or $2k_F$%
), which are smaller than $\Lambda_{\mathrm{f}}$. Accordingly, we introduce
two ``bosonic'' cutoffs: $\Lambda _{\mathrm{b}}$, set by the interaction
with the momentum transfer near $2k_{F}$ ($g_{1}$ vertex),  and ${\bar{%
\Lambda}}_{\mathrm{b}}$, set by the interaction with a small momentum
transfer ($g_{2}$ and $g_{4}$ vertices), and request that both are smaller
than $\Lambda_{\mathrm{f}}$.
More precisely, we assume that
\begin{equation}
\Lambda_{\mathrm{f}}-\Lambda_{\mathrm{b}},\Lambda_{\mathrm{f}}-{\bar \Lambda}%
_{\mathrm{b}}\gg T.  \label{criterion}
\end{equation}
The model interaction is shown in Fig.~\ref{fig:potential}.  We will see
that there is an
interesting dependence of the specific heat on the ratio $\Lambda_{\mathrm{b}%
}/\Lambda_{\mathrm{f}}$,  but no dependence on the ratio ${\bar \Lambda}_{%
\mathrm{b}}/\Lambda_{\mathrm{f}}$.

The two-cutoff model with $\Lambda_{\mathrm{b}} < \Lambda_{\mathrm{f}}$ has
been used in the  canonical 1D bosonization approach, and in the subsequent
analysis of the sine-Gordon model. It was also considered in Refs.%
\onlinecite{twocutoffs,grest} in the analysis of the electron-phonon
interaction in 1D.  We note in passing that, to our knowledge, it has not
been  explicitly verified that the specific heat for the effective
low-energy  model is the same as for the original model of fermions with
parabolic-type dispersion and a generic interaction $U(q)$, i.e.,  that all
contributions to $C(T)$ from fermionic energies exceeding $\Lambda_{\mathrm{f%
}}$ can be absorbed into the  three couplings $g_i$.  We also note that, in
the bosonization procedure invented by AE, which is not based on the $g-$%
ology model,  the cutoff imposed by the interaction is less
restrictive
that the fermionic cutoff (i.e., $\Lambda _{\mathrm{b}}\gg\Lambda _{\mathrm{f%
}}$), because in their theory the propagators of long-wavelength bosonic
modes are obtained by integrating independently over fermionic momenta
linked by the interaction. AE, however, only focused on the truly low-energy
terms with the running coupling, which should not depend on the ratio $%
\Lambda_{\mathrm{b}}/\Lambda_{\mathrm{f}}$. 

Vertices $g_{1}$ and $g_{2}$ in the $g$ -ology model are related to the spin
and charge components of the backscattering amplitude
\begin{equation}
f_{\alpha \beta ,\gamma \delta }(\pi )=f_{c}(\pi )\delta _{\alpha \beta
}\delta _{\gamma \delta }+f_{s}(\pi ){\vec{\sigma}}_{\alpha \beta }\cdot{%
\vec{\sigma}}_{\gamma \delta },  \label{5}
\end{equation}
as $f_{c}(\pi )=2g_{2}-g_{1}$, $f_{s}(\pi )=-g_{1}$. Vertex $g_{4}$ is
related to the forward scattering amplitude $f(0)$ as $g_4=f_{c}(0)-f_{s}(0)$%
.  For generality, we  extend the model from the
$SU(2)$ symmetric to anisotropic case, i.e., assume that all three vertex
functions $g_{i}$ ($i=1,2,4$) have different values $g_{i\parallel }$ and $%
g_{i\perp }$, depending on whether the spins of the fermions in the initial
state are parallel or opposite. For anisotropic case, the spin component of
the backscattering amplitude splits into the
longitudinal and transverse  parts, and we have
\begin{eqnarray}
f_{c}(\pi ) &=&g_{2\parallel } - g_{1\parallel } +g_{2\perp },  \notag \\
f_{s\parallel }(\pi ) &=&g_{2\parallel } - g_{1\parallel } -g_{2\perp },
\notag \\
f_{s\perp }(\pi ) &=&-g_{1\perp }.  \label{6}
\end{eqnarray}

The forward scattering vertex $g_4$ is invariant under RG renormalization,
but the
backscattering vertices $g_1$ and $g_2$ flow \cite{solyom,emery}.
Keeping only the processes with momentum transfers in narrow windows near
either zero or $2k_F$ (the windows are much smaller than  the cutoffs $%
\Lambda_{\mathrm{b}}$ and ${|bar \Lambda}_b$), we have 
\begin{eqnarray}
&&\frac{dg_{1\parallel}}{dL} = \beta_{1\parallel}; ~~\frac{dg_{1\perp}}{dL}
= \beta_{1\perp}  \notag \\
&&\frac{dg_{2\parallel}}{dL} = \beta_{2\parallel}; ~~\frac{dg_{2\perp}}{dL}
= \beta_{2\perp}  \label{2}
\end{eqnarray}
where $L = \log E_F/E$, $E$ is the running energy, and $\beta$ functions
depend on the couplings $g_{1\parallel,\perp}$ and $g_{2\parallel,\perp}$.
In the one-loop approximation,
\begin{eqnarray}
&& \beta_{1\parallel} = -(g^2_{1\parallel} + g^2_{1\perp}), ~~
\beta_{1\perp}= -2 g_{1\perp} \left(g_{1\parallel} - g_{2\parallel} +
g_{2\perp}\right),  \notag \\
&&\beta_{2\parallel} = - g^2_{1\parallel}, ~~ \beta_{2\perp} = -
g^2_{1\perp}.  \label{2_1}
\end{eqnarray}
Re-expressing the couplings in terms of spin and charge components of the
backscattering amplitude, we find that the spin amplitudes $f_{s_\parallel}$
and $f_{s\perp}$ flow to zero under the RG transformation.  The charge
component of the backscattering amplitude $f_{c}(\pi )=(g_{2\parallel
}+g_{2\perp })-g_{1\parallel }$, however,  does not change under the RG flow.

For the SU(2) symmetric case, $\beta_1 = -2 g^2_1/(1 - g_1),~\beta_2 =
\beta_1/2$, and $g_1 (L)$ renormalizes to zero as $1/L$, while $g_2 (L)$
tends to a constant value of a half of the charge amplitude, which is
invariant under RG.

As we said earlier, the key interest of our analysis is to understand  at
which order within the $g-$ology model the running couplings appear in the
specific heat, and what are the contributions to the specific heat which
contain bare couplings.

\subsection{Results}

We first catalog our main results, and then present calculations in the bulk
of the paper.
We computed $C(T)$ in a direct perturbation theory, expanding in powers of
the couplings $g_i$ to order $g^3$. To first two orders in $g_i$, we found
that the specific heat is expressed via bare couplings $g_2$ and $g_4$, and
the effective backscattering coupling $g_1$.
 At third order, we found an extra contribution to $C(T)$, which comes from
low-energies and contains a  cube of the  running backscattering amplitude
on the scale of $T$. Explicitly,  for the anisotropic case, we found for $T
\ll \Lambda_{\mathrm{b}}$  and neglecting $O(g^3)$ contributions with
non-running couplings
\begin{widetext}
\beq
C(T) = \frac{2\pi T}{3v_F}~\left[1 + \left(
{\tilde{g}}_{1\parallel
}-g_{4\parallel }\right) +  \left(
{\tilde{g}}_{1\parallel
}-g_{4\parallel }\right)^2 +
g^2_{4\perp} +
 \frac{1}{2} \left((g^2_{2\parallel} + g^2_{2\perp}) - 2
g_{2\parallel}{\tilde g}_{1\parallel} + ({\tilde g}^2_{1\parallel} +
{\tilde g}^2_{1\perp}) \right)  + 3 {\tilde g}^2_{1\perp} (T) {\tilde g}_{1\parallel} (T) + ... \right], \label{12} \eeq
\end{widetext}
Here $g_4$ and $g_2$ are the
bare couplings, and ${\tilde g}_{1\parallel}$ and ${\tilde g}_{1\perp}$ are
 the
effective couplings on the scale $\Lambda_{\mathrm{b}}$.  The couplings ${%
\tilde g}_{1\parallel} (T)$ and ${\tilde g}_{1\perp} (T)$ are running
couplings on the scale of $T$ -- these are the solutions of the full RG
equations, Eq. (\ref{2}), with effective ${\tilde g}_{1\parallel}$ and ${%
\tilde g}_{1\perp}$  serving as inputs.
\begin{figure}[tbp]
\begin{center}
\epsfxsize=1.0\columnwidth \epsffile{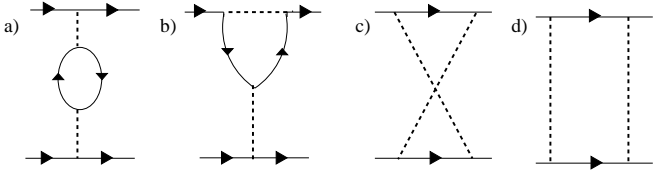}
\end{center}
\caption{One-loop diagrams for the interaction vertices.
In the RG regime (external momenta are much smaller than 
 $\Lambda_b$), the renormalizations of $g_{1\perp}$ and $g_{1\parallel}$ are given by  diagrams
  a), b), and d), while 
 the renormalizations of $g_{2\perp}$ and $g_{2\parallel}$ are given by diagrams c) and d).
  All diagrams give rise to $\log T$ terms.
 For external momenta of  order $\Lambda_b$, only diagram a) gives rise to the 
logarithmic term $\log\left(\Lambda_{{\rm f}}/\Lambda_{{\rm b}}\right)$, as the two fermions
 in the particle-hole bubble can have  momenta
  in the whole range between $\Lambda_{{\rm b}}$ and $\Lambda_{{\rm f}}$.
   For all other diagrams, the interaction constraints the
 internal momenta to be of the same order  as the external momentum,
   and there is no momentum space for the logarithm.}
\label{fig:rg}
\end{figure}

The effective couplings ${\tilde g}_{1\parallel}$ and
${\tilde g}_{1\perp}$ differ from bare $g_{1\parallel,\perp}$ due
to RPA-type
 renormalizations
by $2k_F$ particle-hole bubbles made of fermions  with momenta between $%
\Lambda_{\mathrm{b}}$ and $\Lambda_{\mathrm{f}}$. 
This renormalization comes from diagram a) in Fig.~\ref{fig:rg}).
  There are no such
renormalizations for $g_2$ couplings, which retain their bare
values. We obtain
\begin{subequations}
\begin{eqnarray}
&& {\tilde g}_{1\parallel }\! =\! g_{1\parallel }\! -\! \left(g^2_{1\parallel } +
g^2_{1\perp}\right) L_{\rm{b}}\! +\! \left(g^3_{1\parallel }\! +\! 3 g_{1\parallel }
g^2_{1\perp}\right) L^2_b
 \label{8a} \\
&& {\tilde g}^2_{1\parallel} + {\tilde g}^2_{1\perp}=  g^2_{1\parallel } +
g^2_{1\perp} - 2 \left(g^3_{1\parallel } + 3 g_{1\parallel }
g^2_{1\perp}\right) L_{\rm{b}}
\label{8b}
\end{eqnarray}
\end{subequations}
where $L_{\mathrm{b}} = \log{\left(\Lambda_{\mathrm{f}}
/\Lambda_{\mathrm{b}}\right)}$.

We emphasize that
an RPA-type renormalization is not equivalent to RG,
so that  effective
 ${\tilde g}_{1\parallel}$ and ${\tilde g}_{1\perp}$ differ from the
solutions of (\ref{2}), (\ref{2_1}) already at one-loop order.
 The difference is due to the fact that in
 the  one-loop RG equations, the RPA and ladder-type 
 renormalizations of $g_1$, and 
 the ladder renormalizations of $g_2$ 
    are all coupled, while only the RPA diagrams lead to $L_{\rm{b}}$ terms in the  renormalization
    from $g_1$ to ${\tilde g}_1$ .

 Coupling between the RPA and ladder renormalizations in the RG regime
is absent for the isotropic, $SU(2)$ symmetric case. Then
 ${\tilde g}_1 = g_1/(1 + 2 g_1 L_{%
\mathrm{b}})$ becomes equivalent to one-loop RG. Furthermore, in the symmetric case,  the running $g_1 (L)$ at the lowest energies
 behaves  in the 
 one-loop approximation   as ${\tilde g}_1/(1 + 2 {\tilde g}_1 L)$.
 For the specific
heat we then obtain
\begin{widetext}
\begin{equation}
C(T)=\frac{2\pi T}{3v_{F}}~\left[ 1+  \left(
{\tilde{g}}_{1}-g_{4}\right) +  \left(
{\tilde{g}}_{1}-g_{4}\right)^2  + g^2_4
+ \left(g_{2} -\frac{1}{2}{\tilde{g}}_{1}\right)^2
+ \frac{3}{4} \tilde{g}_{1}^{2}
 +\frac{3{\tilde{g}}_{1}^{3}}{(1+2{\tilde{g}}_{1}L)^{3}}\right]  \label{14}
\end{equation}
\end{widetext}

The results of our direct perturbative analysis are in agreement with the
results  for the Kondo problem \cite{kondo}and XXZ spin chain \cite{lukyanov}
-- for both models, the specific heat was shown to behave as $T/\log ^{3}T$
at the lowest temperatures. These two models are argued to be in the same
universality class as the model of interacting electrons with the
interaction in the spin sector. The same behavior was found by Cardy \cite
{cardy} and Ludwig and Cardy~\cite{ludwig_cardy} in their study of a
conformally invariant theory perturbed by the marginal perturbation from the
fixed point (the sine-Gordon model belongs to this class of theories), and
by AE in their ``multidimensional bosonization'' analysis. In all these
theories, the focus was on the universal terms which are confined to low
energies i.e., are not anomalies. If only such terms are included, the full
spin contribution to the specific heat scales as $T/\log ^{3}T$ in the $SU(2)
$ isotropic case, i.e., the spin part of the specific heat coefficient
vanishes at $T=0$. Our direct perturbation theory reproduces the same
universal behavior in the spin sector, but also generates extra
contributions to the specific heat which contain effective interaction on
the scale of $\Lambda_{\mathrm{b}}$.

To make the comparison with the bosonization and sine-Gordon model explicit,
we re-write our result via spin and charge velocities $v_{F}u_{\rho }$ and $%
v_{F}u_{\sigma }$ obtained by diagonalizing the gradient part of the
Hamiltonian:
\begin{eqnarray}
&&u_{\rho }^{2}\!=\!(1+g_{4\parallel }+g_{4\perp }-g_{1\parallel
})^{2}\!-\!(g_{2\parallel }+g_{2\perp }-g_{1\parallel })^{2},  \notag \\
&&u_{\sigma }^{2}\!=\!(1+g_{4\parallel }-g_{4\perp }-g_{1\parallel
})^{2}\!-\!(g_{2\parallel }-g_{2\perp }-g_{1\parallel })^{2}  \label{15}
\end{eqnarray}
Using (\ref{15}), one can re-write (\ref{12}) as
\begin{widetext}
\beq C(T) = \frac{\pi T}{3v_F} \left(\frac{1}{{\tilde u}_{\rho}} +
\frac{1}{{\tilde u}_{\sigma}}\right) + \frac{\pi T}{3 v_F} {\tilde
g}_{1\perp}^2 + \frac{2 \pi T}{v_F}~ g^2_{1\perp} (T) g_{1\parallel}
(T). \label{16} \eeq
\end{widetext}
 The last term in (\ref{16}) is the universal contribution from low energies.
The first term is the sum of the specific heats of two gases of
free particles with the effective velocities ${\tilde{u}}_{\rho }$
and ${\tilde{u}}_{\sigma },$ which are the same as in (\ref{15})
except for $g_{1||}$ and $g_{1\perp } $ are now the effective,
renormalized vertices. The term in the middle is  an additional
contribution from the spin channel. Very likely, this contribution
can be absorbed into the renormalization of  spin velocity
${\tilde u}_\sigma \rightarrow {\tilde u}_\sigma - {\tilde
g}_{1\perp}^2$, i.e., the specific heat can be re-expressed  as
the sum of the contribution with running couplings, and the
specific heat of two ideal gases of fermions with bare charge
velocity (albeit with ${\tilde g}_1$), and the renormalized spin
velocity.

In the rest of the paper, we present the details of our calculations. In
Sec. \ref{1st}, we outline the computational procedure, calculate the
first-order diagram for the thermodynamic potential, and demonstrate
explicitly the sensitivity of the result for $C(T)$ to the ratio of the
cutoffs.  In Sec.\ref{2nd} and \ref{3rd}, we compute second and third order
diagrams for the thermodynamic potential,  and discuss the fourth-order
result.  In Sec.\ref{sec:sine}, we analyze the specific heat in the
framework of the sine-Gordon model. Sec. \ref{conclusions} presents the
conclusions. Some technical details of the calculations are presented in the
Appendices.

\section{Perturbation theory and the role of cutoffs}

\label{cutoff}

\subsection{Preliminaries}

\label{prelim}

In this and the next two Sections we set $v_{F}=1$. We restore $v_{F}$ in
the final formulas for the specific heat.

The specific heat of an interacting system of fermions can be extracted from
the thermodynamic potential $\Xi $ via $C(T)=-T\partial ^{2}\Xi /\partial
T^{2}$. The thermodynamic potential is given by the Luttinger-Ward formula:
\begin{equation}
\Xi =\Xi ^{(0)}-2T\sum_{\omega }\int \frac{dk}{2\pi }\left[ \log \left(
G_{0}G^{-1}\right) -\Sigma G+\sum_{\nu }\frac{1}{2\nu }\Sigma _{\nu }G\right]
\label{17}
\end{equation}
where
\begin{equation}
\Xi ^{(0)}=-2T\sum_{\omega }\int \frac{dk}{2\pi }\left[ \frac{1}{2}\log
\left( \epsilon _{\mathbf{k}}^{2}+\omega _{m}^{2}\right) \right],  \label{18}
\end{equation}
is the thermodynamic potential of the free Fermi gas per unit length $%
G_{0}=\left( i\omega _{m}-\epsilon _{k}\right) ^{-1}$, $\epsilon _{k}$ is
the dispersion, $G=\left( i\omega _{m}-\epsilon _{k}+\Sigma \right) ^{-1}$, $%
\Sigma $ is the exact (to all orders in the interaction) self-energy, and $%
\Sigma _{\nu }$ is the skeleton self-energy of order $\nu $. The skeleton
and full self-energy are related via $\Sigma =\sum_{\nu }\Sigma _{\nu }$ and
are evaluated at finite $T$. Expanding both $G$ and $\Sigma _{\nu }$ in Eq.~(%
\ref{16}) in powers of the interaction, one generates a perturbative
expansion for $\Xi $ in the series of closed diagrams with no external legs.

The free-fermion expression for $C(T)$ is obtained from Eq. (\ref{18}). At
low $T$, the momentum integration is confined to $k \approx \pm k_F$ and
yields

\begin{equation}
\Xi^{(0)}=- T\sum_{\omega _{m}}|\omega _{m}| = - \frac{\pi T^2}{3} + {\text
const},  \label{19}
\end{equation}

such that $C^{(0)} (T) = 2\pi T/3$.

\subsection{First order diagrams}

\label{1st}
\begin{figure}[tbp]
\begin{center}
\epsfxsize=1.0\columnwidth \epsffile{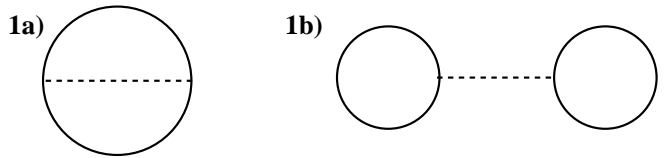}
\end{center}
\caption{First order diagrams for the thermodynamic potential. Here and in
the rest of the figures, the dashed line represents the interaction. Diagram
1b) does not contribute to the temperature dependence of the thermodynamic
potential.}
\label{fig:first}
\end{figure}
At first order, the $T$ dependence of $\Xi $ comes  from the bubble diagram
crossed by the interaction  line [diagram 1a) in Fig.\ref{fig:first}).  This
diagram contains two contributions: one with a small momentum transfer and
another with a momentum transfer near $2k_{F}$. As spin is conserved along
the bubble, the corresponding coupling constants are $g_{4\parallel }$ and $%
g_{1\parallel },$ respectively.

The safe way to evaluate the diagram is to sum over  frequencies first, as
the frequency summation is constrained neither by the interaction nor by the
fermionic bandwidth, and then integrate over the
 fermionic momentum $k$ and
bosonic, transferred momentum $q$.
We will measure $q$ as a
deviation from zero for $g_{4\parallel }$ term, and from $2k_F$
for $g_{1\parallel }$ term, and, as we said,  will cut
interactions at $|q| = {\bar{\Lambda}}_{\mathrm{b}}$ for forward
scattering process and at $\Lambda _{\mathrm{b}}$ for
backscattering process. We linearize the fermionic dispersion near
the Fermi surface and set the cutoff
of the integration over $k$ at $|k \pm q/2| \leq \Lambda _{\mathrm{f}}$, $%
\Lambda _{\mathrm{f}} > {\bar{\Lambda}}_{\mathrm{b}}, \Lambda _{\mathrm{b}}$.

For small momentum transfer, we then obtain
\begin{equation}
\Xi ^{(1)}_{q=0}= \frac{2 g_{4\parallel }}{\pi }\int_{0}^{{\bar \Lambda} _{%
\mathrm{b}}}\!\!\!\! dq\int_0^{\Lambda_{\mathrm{f}} -q/2}\!\!\!\! dk \frac{%
\cosh {\frac{q}{2T}}}{\cosh {\frac{q}{2T}}+\cosh {\frac{k}{T}}},  \label{20}
\end{equation}
and for the momentum transfer near $2k_{F}$, we obtain
\begin{equation}
\Xi ^{(1)}_{2k_{F}}=\frac{2 g_{1\parallel }}{\pi }\int_0^{\Lambda_{\mathrm{b}%
}}\!\!\!\! dq \int_0^{\Lambda_{\mathrm{f}} -q/2} \!\!\!\! dk \frac{\cosh {%
\frac{k}{T}}}{\cosh {\frac{q}{2T}}+\cosh {\frac{k}{T}}}.  \label{21}
\end{equation}
Subtracting $T$-independent terms in (\ref{20}) and (\ref{21}) and
introducing rescaled variables $x=k/T$, $y=q/(2T)$, we re-write (\ref{20})
and (\ref{21}) as
\begin{eqnarray}
\Xi ^{(1)}_{q=0}&=&\frac{2 g_{4\parallel }T^{2}}{\pi } \int_{0}^{{\bar
\Lambda} _{\mathrm{b}}/2T}\!\!\!\! dy\int_0^{\frac{\Lambda_{\mathrm{f}}}{T}
-y}\!\!\!\! dx \frac{\cosh {y}-\cosh {x}}{\cosh {y}+\cosh {x}},  \label{22}
\end{eqnarray}
and
\begin{equation}
\Xi ^{(1)}_{2k_{F}}=\frac{2g_{1\parallel }T^{2}}{\pi } \int_0^{\Lambda_{%
\mathrm{b}}/2T}\!\!\!\! dy \int_0^{\frac{\Lambda_{\mathrm{f}}}{T} -y}
\!\!\!\!dx \frac{\cosh {x}-\cosh {y}}{\cosh {x}+\cosh {y}}.  \label{23}
\end{equation}
We immediately see that the first-order contribution to the
thermodynamic potential vanishes if we formally extend the
integrals over $x$ and $y$ to infinity. Integrals (\ref{22}) and
(\ref{23}) are  similar to the integrals which give rise to
anomalies in the field theory \cite{jackiw}. The integrands are
odd under the interchange of $x$ and $y$; therefore universal,
cutoff independent contributions apparently vanish, but the 2D
integrals are ultraviolet divergent, if we set $T$ to zero. A
finite $T$ then sets an ultraviolet regularization of the
divergent 2D integral and gives rise to finite terms in $\Xi $
which do not explicitly depend on the cutoff. By analogy with the
field theory, hereafter we refer to these terms as ``anomalies''.

For definiteness, we focus on the $2k_{F}$ contribution. Since $\Lambda _{%
\mathrm{f}}>\Lambda _{\mathrm{b}}$
(in the sense of Eq.(\ref{criterion}),
the integration over $y$ extends to a much narrower range than that over $x$%
. In this situation, the most natural way to evaluate the thermal part of $%
\Xi ^{(1)}_{2k_{F}}$ is to re-express (\ref{23}) as
\begin{eqnarray}
&&\Xi ^{(1)}_{2k_{F}}=-\frac{4}{\pi }g_{1\parallel }T^{2}\int_{0}^{\Lambda _{%
\mathrm{b}}/2T}dy\cosh {y}  \notag \\
&&\times \int_{0}^{\frac{\Lambda _{\mathrm{f}}}{T} -y}\frac{dx}{\cosh {x}%
+\cosh {y}}.  \label{23_1}
\end{eqnarray}
The integral over $x$ now converges, and, because $\Lambda _{\mathrm{f}%
}>\Lambda _{\mathrm{b}}$, we can safely set the upper limit of the $x$
integral to infinity. The $x$ integration then can be performed exactly and
yields
\begin{eqnarray}
&&\Xi ^{(1)}_{2k_{F}}=-\frac{4}{\pi }~g_{1\parallel }T^{2}\int_{0}^{\Lambda
_{\mathrm{b}}/2T}dyy\coth {y},  \notag \\
&=&-\frac{g_{1\parallel }}{2\pi }\Lambda _{b}^{2}-\frac{4g_{1\parallel }}{%
\pi }T^{2}\int_{0}^{\Lambda _{\mathrm{b}}/2T}dy y(\coth {y}-1).  \label{23_2}
\end{eqnarray}
The thermal part of $\Xi ^{(1)}_{2k_{F}}$ comes from the second term
\begin{equation}
\Xi ^{(1)}_{2k_{F}}=\mathrm{const}-\frac{\pi }{3}g_{1\parallel }T^{2}.
\label{23_3}
\end{equation}
Observe that the $T^{2}$ piece is independent of the cutoff. Furthermore, in
this computational procedure, the frequency sums and the momentum integrals are
fully ultraviolet convergent, and Eq. (\ref{23_3}) comes from small momenta $%
k,q\sim T\ll \Lambda _{\mathrm{f}},$ $\Lambda _{\mathrm{b}}$.

Alternatively, however, we can  evaluate the integrals in (\ref{23}) by
integrating over the (dimensionless)
bosonic momentum $y$ first. To do this, we  neglect $y$ in the upper limit
of the integral over $x$ (we will check a'posteriori that this is
justified), and re-express (\ref{23}) as
\begin{eqnarray}
&&\Xi ^{(1)}_{2k_{F}}=\frac{4g_{1\parallel }}{\pi }T^{2}\int_{0}^{\Lambda _{%
\mathrm{f}}/T}dx\cosh {x}  \notag \\
&&\times \int_{0}^{\Lambda _{\mathrm{b}}/2T}\frac{dy}{\cosh {x}+\cosh {y}}.
\label{23_4}
\end{eqnarray}
It is tempting to set the upper limit of the $y$ integral to infinity, as
this integral converges. However, one has to be cautious as there is a range
of $x$ where $\cosh {x}>\cosh {y}$ for any $y$. To see how this affects the
result, we represent the $y$ integral as
\begin{widetext}
\beq
 \int^{\Lambda_{\mathrm{b}}/2T}_{0} \frac{d y} {\cosh{x} + \cosh{y}} =
\int^{\infty}_{0} \frac{d y} {\cosh{x} + \cosh{y}} -
\int^{\infty}_{\Lambda_{\mathrm{b}}/2T} \frac{d y} {\cosh{x} + \cosh{y}}  =
\frac{x}{\sinh{x}} -  \frac{1}{\cosh{x}}
\int^{\infty}_{\left(\Lambda_{\mathrm{b}}/2T\right)} \frac{d y} {1 + e^{y-x}}.
\label{23_5} \eeq
\end{widetext}
We replaced $\cosh {x}$ and $\cosh {y}$ in the last term by the
exponentials, as $y$ are large, and we anticipate typical $x$ to be large as
well. The remaining integration is straightforward, and we obtain
\begin{widetext}
\beq \Xi^{(1)}_{2k_F} = {\rm const} +
 \frac{\pi}{3} g_{1\parallel} T^2 -
\frac{4 g_{1\parallel}}{\pi} T^2 \int^{\Lambda_{\mathrm{f}}/T}_0
 d x  \log \left(1 + e^{x- \Lambda_{\mathrm{b}}/2T}\right).
\label{23_6}
\eeq
\end{widetext}
The first term is the contribution from low energies -- the same as in (\ref
{23_3}), but with the opposite sign. The second term by construction is  the
contribution from energies much larger than $T$.  Evaluating the second
integral, we find  that it also contributes
 a $T^2$  term
to $\Xi^{(1)}_{2k_F}$:
\begin{eqnarray}
&&\frac{4g_{1\parallel }}{\pi }T^{2}\int_{0}^{\Lambda _{\mathrm{f}}/T}dx\log
\left( 1+e^{x-\Lambda _{\mathrm{b}}/2T}\right)  \notag \\
&=&\mathrm{const}+\frac{2\pi }{3}g_{1\parallel }T^{2} ,  \label{23_7}
\end{eqnarray}
The $T^2$ term in (\ref{23_7}) comes from $x \sim
\Lambda_{\mathrm{b}}/2T$. It is essential that these $x$ are smaller than
the upper limit of $x-$integration,  otherwise such a contribution would not
exist.  Substituting this back into (\ref{23_6}), we find that the
high-energy term is opposite in sign and twice larger than the low-energy
one, so that the sum of the two contributions is given precisely by Eq.(\ref
{23_3}). Going back through the derivation of (\ref{23_7}), we  see that
typical $y$ and $x$ are near $\Lambda_{\mathrm{b}}/2T$, well below the upper
limit of the $x$ integration. In this situation, the neglect of $y$ in the
upper limit of the integral over $x$ is legitimate, to accuracy $\exp{%
-\Lambda_{\mathrm{b}}/T}$.

We see therefore that $\Xi ^{(1)}_{2k_{F}}$ can be equally
 well obtained either
as a low-energy contribution or as a high-energy one. This is a hallmark of
an anomaly. The same is true also for the forward scattering term $\Xi
^{(1)}_{q=0}$: the $T^{2}$ term again can be equally obtained as a
low-energy contribution or as a contributions from energies of order ${\bar
\Lambda}_{\mathrm{b}}$.

Combining the results for backscattering and forward scattering, we obtain
for the specific heat
\begin{equation}
C^{(1)}(T)=\frac{2\pi T}{3v_{F}}\left( ~g_{1\parallel } -g_{4\parallel
}\right).  \label{24}
\end{equation}

As we said in the Introduction, the $g_{1\parallel }$ term in (\ref{24}) is
not present in the standard  bosonization approach~\cite{giamarchi_book}.
The physical argument  is that the $g_{1\parallel}$ should only appear in $%
C(T)$ in the combination $g_{1\parallel} - g_{2\parallel}$ as the two
vertices transform into each other  by interchanging external momenta
without interchanging spins, and therefore are physically indistinguishable
\cite{schulz_95,metzner}. The first order diagram with $g_{2\parallel}$ is a
Hartree diagram with two bubbles connected by the interaction at exactly
zero transferred momentum [diagram 1b) in Fig.\ref{fig:first}]. As each of
these two bubbles represents a total electron density, this diagram
obviously does not depend on $T$. By the argument above, the diagram with $%
g_{1\parallel}$ also should not depend on $T$. This consideration is,
however, only valid if the cutoffs are infinite. For finite cutoffs, there
appears an extra  ``anomaly-type'' contribution, in which $g_{1\parallel}$
appears in the combination with $g_{2\parallel}$, as we just demonstrated~
\cite{kostya_private}.  A similar reasoning within real-space consideration
has been presented in \cite{capponi}. Another argument for presence of the $%
g_{1\parallel }$ term is based on the observation that fermions with the
same spin do not interact via a contact interaction; hence the interaction
should drop out of the results in this limit. This implies that the
observables, such as $C(T)$ must depend separately on the combinations $%
g_{4||}-g_{1||}$ and $g_{2||}-g_{1||}$~ \cite{starykh_maslov,matveev}.  Eq. (%
\ref{24}) is consistent with this argument as the limit of a contact
interaction,
i.e, for $g_{1\parallel }=g_{4\parallel }$, $C^{(1)}(T)$ vanishes.

The interplay between low-energy and high-energy contributions to $\Xi $ can
be also understood if one interchanges one momentum integration and one
frequency  summation and expresses $\Xi^{(1)}$ via the polarization bubble
as
\begin{eqnarray}
&&\Xi ^{(1)}_{q=0}=-g_{4\parallel }T\sum_{\Omega }\int dq\Pi _{q=0}(q,\Omega
),  \notag \\
&&\Xi ^{(1)}_{2k_{F}}=-g_{1\parallel }T\sum_{\Omega }\int_{-\Lambda _{%
\mathrm{b}}}^{\Lambda _{\mathrm{b}}}dq\Pi _{2k_{F}}(q,\Omega ).  \label{25}
\end{eqnarray}
The sub-indices indicate that the momentum integration is confined to $q$
near zero or near $2k_{F}$.

For briefness, we consider only the backscattering term. The polarization
bubble $\Pi_{2k_{F}}(q,\Omega )$ is given by
\begin{widetext}
\bea \Pi_{2k_F} (q, \Omega)&\equiv&
2T\sum_{\omega}\int \frac{dk}{2\pi}G_{R}(\omega+\Omega,k+q)G_{L}(\omega,k) +
G_{L}(\omega+\Omega,k+q)G_{R}(\omega,k) \\
&& = \frac{1}{2\pi} \left(\log{\frac{\Omega^2 +q^2}{4
\Lambda^2_f}} - 8 \int_0^\infty dk k n_F (k)
\left(\frac{1}{(q-i\Omega)^2 - 4 k^2} + \frac{1}{(q + i \Omega)^2
- 4 k^2}\right)\right), \label{26} \eea
\end{widetext}
where $G_{L,R}(k\omega )$ is the Green's function of right/left moving
fermions and $n_{F}(x)$ is the Fermi function. The first term in Eq.(\ref{26}%
) is the zero-temperature Kohn anomaly, the rest is the thermal
contribution. The integration over $k$ gives the result for $\Pi(q,\Omega)$
in terms of di-Gamma functions~\cite{schulz}, but for our purposes it is
more convenient to use (\ref{26}).

Substituting (\ref{26}) into (\ref{25}), we find
\begin{equation}
\Xi ^{(1)}_{2k_{F}}=- g_{1\parallel} ~(Q +P),
\end{equation}
where $Q$ and $P$ are the contributions from the Kohn anomaly and from the
thermal piece in (\ref{26}), respectively. The temperature
 -dependent part of the $Q$ term is
\begin{equation}
Q= \frac{1}{2\pi}~ T\sum_{\Omega }\int_{-\Lambda _{\mathrm{b}}}^{\Lambda _{%
\mathrm{b}}}dq\log {\frac{\Omega ^{2}+q^{2}}{4\Lambda^2_f}},  \label{27_1}
\end{equation}
comes from low energies regardless of whether the sum or the integral is
done first. In both cases, we get, up to a constant,
\begin{equation}
Q=Q_L = -\pi T^{2}/3,  \label{27_2}
\end{equation}
where subindex L specifies that this is a contribution from low energies: $%
\Omega, q \sim T$. The second, thermal, term is determined either by low or
by high energies, depending on the order. If the momentum integration is
done first, the non-zero result is obtained only because $\Lambda _{\mathrm{b%
}}$ is finite; otherwise, the integration contour can be closed in that
half-plane where the integrand has no poles. Re-arranging the integrals, we
rewrite this  contribution as
\begin{widetext}
\bea
&&P = - \frac{4}{\pi} T \sum_\Omega \int^{\Lambda_{\mathrm{b}}}_{-\Lambda_{\mathrm{b}}} dq
\int_0^\infty dk k n_F (k) \left(\frac{1}{(q-i\Omega)^2 - 4 k^2} + \frac{1}{(q + i \Omega)^2 - 4 k^2}\right), \nonumber\\
&&  =\frac{8}{\pi} T \sum_\Omega \int^{\infty}_{\Lambda_{\mathrm{b}}} dq
\int_0^\infty dk k n_F (k)   \left(\frac{1}{(q-i\Omega)^2 - 4 k^2} +
\frac{1}{(q + i \Omega)^2 - 4 k^2}\right). \label{28} \eea
\end{widetext}
As the Fermi function in (\ref{26}) confines the fermionic momentum to $%
k\sim T$, and $q>\Lambda _{\mathrm{b}}$ is large, we can neglect $4k^{2}$
compared to $(q\pm i\Omega )^{2}$ in the denominator. This simplifies $P$ to
\begin{equation}
P=\frac{2\pi T^{2}}{3}T\sum_{\Omega }\int_{\Lambda _{\mathrm{b}}}^{\infty
}dq\left( \frac{1}{(q-i\Omega )^{2}}+\frac{1}{(q+i\Omega )^{2}}\right)
\label{29}
\end{equation}
Performing the momentum integration, we obtain
\begin{equation}
P=\frac{4\pi T^{2}}{3}T\sum_{\Omega }\frac{\Lambda _{\mathrm{b}}}{\Lambda _{%
\mathrm{b}}^{2}+\Omega ^{2}}.  \label{30}
\end{equation}
Evaluating the integral, we find that $P$ does not depend on the cutoff, and
equals to
\begin{equation}
P = P_{H}=\frac{2\pi T^{2}}{3},  \label{31}
\end{equation}
where subindex $H$ specifies that this is a contribution from high energies $%
\Omega, q \sim \Lambda _{\mathrm{b}}$.

Alternatively, $P$ can be evaluated by doing frequency summation first. One
can easily check that, to order $T^{2}$, the frequency sum can be replaced
by the integral. The frequency integral is non-zero only for $q<2k$,
otherwise the poles in $\Omega $ are located in the same half-plane, and the
frequency integral vanishes. Evaluating the frequency integral and then the
integral over $q$, we reduce $P$ to
\begin{equation}
P = P_{L}=\frac{8}{\pi }\int_{0}^{\infty }dkkn_{F}(k)=\frac{2\pi T^{2}}{3},
\label{32}
\end{equation}
where subindex $L$ specifies that this a contribution from low energies $%
\Omega \sim T$.  We see that $P_{H}=P_{L}$, i.e.,  the same result for $P$
can be obtained either as a high-energy contribution, or as low-energy one.
In both cases $P$ is formally independent of the cutoff, and the total
backscattering part of $\Xi ^{\left( 1\right) }$ is given by
\begin{equation}
\Xi ^{(1)}_{2k_{F}}=\mathrm{const}-g_{1\parallel }(Q+P)=\mathrm{const}-\frac{%
g_{1\parallel }\pi }{3}T^{2},  \label{33}
\end{equation}
which coincides with (\ref{23_3}).

Which of the two ways (low-energy or high-energy) is physically correct? As
we discussed in the Introduction, if $\Xi^{(1)}_{2k_{F}}$ comes from low
energies (of order $T$), one should expect $T\log T$ terms in $C(T)$ already
at the next (second) order; on the contrary, if it comes from high energies,
no such terms are expected. AE suggested implicitly that the correct
procedure is to take the average of two possible orderings, i.e., to
represent frequency summation and momentum integration in (\ref{25}) as
\begin{equation}
\frac{1}{2}\left( T\sum_{\Omega }~\int dq+\int dq~T\sum_{\Omega }\right) .
\label{34}
\end{equation}
In this procedure, $P$ in Eq. (\ref{28}) is a sum
\begin{equation}
P=\frac{1}{2}P_{L}+\frac{1}{2}P_{H},
\end{equation}
with $P_{H}=P_{L}=\pi T^{2}/3$. Total $\Xi ^{(1)}_{2k_{F}}$ is the sum of $P$
and $Q$ [see Eq.(\ref{33})], where $Q = Q_L=-\pi T^{2}/3$ comes from low
energies. Adding $P$ and $Q$, we find that the low-energy contributions
cancel out, and the net result for $\Xi ^{(1)}_{2k_{F}}$ is the high-energy
contribution.

Eq. (\ref{34}), however, contains some ambiguity, as one could equally well
 can re-write
$P$ as $\alpha P_{L}+(1-\alpha )P_{H}$ with an arbitrary coefficient $\alpha
$. Then, the balance between the low and high energy contributions to $\Xi
^{\left( 1\right) }$ would depend on $\alpha$. Whether $\Xi ^{\left(
1\right) }$ contains running or bare coupling $g_{1||}$ can only be
established by an explicit computation to the next (second) order. This is
what we will do in next Section.

\subsection{Second order diagrams}

\label{2nd}
\begin{figure}[tbp]
\begin{center}
\epsfxsize=1.0\columnwidth \epsffile{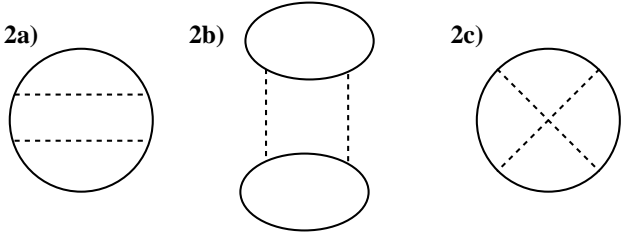}
\end{center}
\caption{Second order diagrams for the thermodynamic potential.}
\label{fig:second}
\end{figure}
The second-order diagrams for the thermodynamic potential are shown in Fig.~%
\ref{fig:second}.  There are  two different types of diagrams, obtained by
inserting either self-energy corrections or vertex corrections into the
first-order diagram. The diagram with self-energy insertions [diagram 2a)]
is readily computed either explicitly, or by evaluating first-order
self-energy and substituting the result into the first-order diagram. We
will not discuss computational steps (they are not qualitatively different
from those to first order) and present only the final result: the diagram
2a) yields a regular $T^{2}$ contribution to $\Xi $ of the form
\begin{equation}
\Xi _{2a}^{(2)}=-\frac{1}{3\pi }T^{2}\left( g_{1\parallel }-g_{4\parallel
}\right) ^{2}.  \label{35}
\end{equation}
Next, there are vertex correction diagrams 2b) and 2c), which involve the
forward scattering vertex $g_{4}$ (small transferred momentum and $2k_{F}$
total incoming momentum) and $g_{2}$ vertex (small transferred and small
total incoming momentum).  The contributions of order $g_{4\parallel }^{2}$,
$g_{4\perp }^{2}$, $g_{2\parallel }^{2}$, $g_{2\perp }^{2}$, and also of
order $g_{2\parallel }g_{1\parallel }$, are all expressed via the bilinear
combinations of the bubbles for right and left movers
\begin{eqnarray}
\Pi _{L,R}(q,\Omega ) &\equiv &T\sum_{\omega }\int \frac{dk}{2\pi }%
G_{R,L}(k+q,\omega +\Omega )G_{R,L}(k,\omega )  \notag \\
&=&\pm \frac{1}{2\pi }\frac{q}{\imath \Omega \mp q},
\end{eqnarray}
whose sum is the total polarization bubble $\Pi _{q=0}(q,\Omega )=\Pi
_{R}(q,\Omega )+\Pi _{L}(q,\Omega ).$ The evaluation of $T\sum_{\Omega }\int
dq\Pi _{i}(q,\Omega )\Pi _{j}(q,\Omega )$ ($i,j=L,R)$ is straightforward,
and the result does not depend on the order of momentum and frequency
integrations. We have, up to $T $-independent terms,
\begin{widetext}
\bea
&&T \sum_\Omega \int d q \left[\Pi_L^2  (q, \Omega)+ \Pi_R^2  (q, \Omega)\right] = - \frac{1}{\pi^2} T \sum_\Omega \int d q \frac{\Omega^2}{\Omega^2 + q^2} = - \frac{1}{\pi}  T \sum_\Omega |\Omega| = {\rm const} + \frac{T^2}{3}, \nonumber\\
&&T \sum_\Omega \int d q \Pi_L  (q, \Omega) \Pi_R (q, \Omega) = -
\frac{1}{4\pi^2}  T \sum_\Omega \int d q \frac{\Omega^2}{\Omega^2
+ q^2} =   {\rm const} + \frac{T^2}{12}. \label{37} \eea
\end{widetext}
These expressions give rise only to regular $T^{2}$ terms in the
thermodynamic potential and, consequently, to $T$ terms in the specific
heat. Collecting combinatorial factors, we find that the contribution from $%
g^2_{4\parallel }$ cancels out among diagrams 2b) and 2c), while
the rest yields
\begin{equation}
\Xi _{\mathrm{reg}}^{(2)}=-\frac{1}{3\pi }T^{2}\left[ g_{4\perp }^{2}+\frac{1%
}{2}\left( g_{2\parallel }^{2}+g_{2\perp }^{2}\right) -g_{2\parallel
}g_{1\parallel }\right].  \label{38}
\end{equation}

Non-trivial second-order contributions are associated with the vertex
corrections due to backscattering amplitude  in diagram 2b). These are most
easily expressed via the square of the $2k_{F}$ polarization bubble as
\begin{equation}
\Xi ^{(2)}_{2k_F}=-\frac{\pi }{2}~T\sum_{\Omega }\int dq\left( g_{1\parallel
}^{2}+g_{1\perp }^{2}\right) \Pi _{2k_{F}}^{2}(q,\Omega ).  \label{39}
\end{equation}
The evaluation of $T\sum_{\Omega }\int dq\Pi _{2k_{F}}^{2}(q,\Omega )$,
presented in Appendix A, gives
\begin{equation}
T\sum_{\Omega }\int dq\Pi _{2k_{F}}^{2}(q,\Omega )=\frac{T^{2}}{3}\left( 1-2
L_{\mathrm{b}} \right),  \label{40}
\end{equation}
where, we remind, $L_{\mathrm{b}} = \log\left(\Lambda _{\mathrm{f}}/\Lambda
_{\mathrm{b}}\right)$.  Combining (\ref{39}) and (\ref{40}), we obtain
\begin{equation}
\Xi ^{(2)}_{2k_F}=-\frac{\pi T^{2}}{6}\left( g_{1\parallel }^{2}+g_{1\perp
}^{2}\right) \left( 1 - 2 L_{\mathrm{b}} \right).  \label{41}
\end{equation}

The logarithmic term in (\ref{41}) is the contribution from particle-hole
bubble with fermions with energies between $\Lambda_{\mathrm{b}}$ and $%
\Lambda_{\mathrm{f}}$. One can easily verify that it coincides with one-loop
renormalization  of the backscattering amplitude $g_{1\parallel }$, which
also comes from the  bubble diagram. Indeed, according to (\ref{8a}), the
effective coupling ${\tilde g}_{1\parallel }$ on the scale $\Lambda_{\mathrm{%
b}}$  is, to order $g^2$,
\begin{equation}
{\tilde g}_{1\parallel } =g_{1\parallel }-(g_{1\parallel }^{2}+g_{1\perp
}^{2})L_{\mathrm{b}}.  \label{42}
\end{equation}
This is precisely what one obtains by combining $\Xi ^{(1)}_{2k_{F}}$ from (%
\ref{33}) and the logarithmic term in $\Xi ^{(2)}_{2k_F}$. We see that at
low $T$, the effective ${\tilde g}_{1\parallel}$  is the renormalized
coupling on the scale of $\Lambda_{\mathrm{b}}$ rather than on the scale of $%
T$. This implies that the correct way to interpret the anomaly in $\Xi
^{(1)}_{2k_{F}}$ is to treat it as a purely high-energy contribution. This
agrees with the ``symmetrized'' procedure of Eq. (\ref{34}).

If the fermionic and bosonic cutoffs differ significantly, i.e., $\Lambda_{%
\mathrm{f}}\gg \Lambda_{\mathrm{b}}$, then $g_{1||}$ flows logarithmically
in the energy interval $\Lambda_{\mathrm{b}}\ll E\ll \Lambda_{\mathrm{f}}$.
This flow freezes, however, at $E\sim \Lambda_{\mathrm{b}}$.
We also emphasize that the non-logarithmic term in $\Xi^{(2)}_{2k_F}$ is
independent of
the ratio of the fermionic and bosonic cutoff, just like first-order $%
g_{1\parallel} - g_{4\parallel}$ term. Another way to see this is to adopt a
different computational procedure for the backscattering part of diagram 2b.
Namely, by virtue of $2k_F$ scattering, two pairs of fermions from different
bubbles have nearly equal momenta. Combining these two pairs into two
bubbles with small momentum transfers and integrating independently over the
two running momenta in these two new bubbles, one can re-express the $%
g_{1}^{2}$ contribution via the product of two polarization bubbles with
small momentum transfers.  This procedure was employed
in earlier work
for $D>1$ \cite{cmgg}, and by AE for
1D. It is justified, however, only when  the momentum dependence of the
interaction is weak up to  the fermionic cutoff, i.e., in a formal limit
when $\Lambda _{\mathrm{b }}\gg \Lambda _{\mathrm{f}}$ (which is opposite to
 what we assume here). Applying this procedure, one can re-express $\Xi^{(2)}_{2k_F}$ as
\begin{widetext}
\beq \Xi^{(2)}(2k_F) =  - 2\pi \left(g^2_{1\parallel} +
g^2_{1\perp}\right) T \sum_\Omega \int d q \Pi_L  (q, \Omega) \Pi_R
(q, \Omega)   = - \frac{\pi T^2}{6}  \left(g^2_{1\parallel} +
g^2_{1\perp}\right). \label{43} \eeq
\end{widetext}
This agrees with Eq. (\ref{41}) without the logarithmic term.

Combing the contributions to $\Xi ^{(2)}$
from Eqs.(\ref{35},\ref{38}) and (\ref{40}),
 we obtain for the second-order specific heat $C^{(2)}(T)$
\begin{widetext}
\begin{equation}
C^{(2)}(T)=-\frac{2\pi T}{3v_{F}}~\left( g_{1\parallel }^{2}+g_{1\perp
}^{2}\right) L_{\rm{b}}
+\frac{2\pi T}{3v_{F}} \left((g_{1\parallel }~
-g_{4\parallel })^{2}  +  g^2_{4\perp}\right) +
+\frac{\pi T}{3v_{F}}\left( (g_{2\parallel } - g_{1\parallel})^{2}
++g_{2\perp}^{2}+g_{1\perp}^{2}\right),  \label{4}
\end{equation}
\end{widetext}

Note that the bilinear combination of $g_{2}$ and $g_{1}$ in the last
term of
(\ref{4}) is precisely the same combination of the
backscattering amplitudes as for the  non-analytic term in $C(T)$ in higher
dimensions (c.f. Eq. (1)):
\begin{eqnarray}
&& (g_{2\parallel } - g_{1\parallel})^{2}+g_{2\perp}^{2}+g_{1\perp}^{2}
\notag \\
&=&\frac{1}{2}\left( f_{c}^{2}(\pi )+f_{s\parallel }^{2}(\pi )+2f_{s\perp
}^{2}(\pi )\right).  \label{11}
\end{eqnarray}

We see that besides  the logarithmic term which transforms $g_{1\parallel }$
into ${\tilde g}_{1\parallel}$, the specific heat  $C^{(2)}(T)$
also contains the
``universal'' second-order terms that
do not depend on the cutoffs. This poses the same question as before --  are
those couplings the running ones (on the scale of $T$)  or the bare ones (on
the scale of a cutoff)? On one hand, the combination of the
second-order $g_{2}$ and $g_{1}$ terms in $C^{(2)}(T)$ is the sum of the
squares of charge and spin components of the backscattering amplitude, Eq. (%
\ref{11}). As the spin amplitude flows under RG and acquires $\log T$
corrections, one could expect $T\log T$ terms at the next, third order. On
the other hand, \textit{all} constant terms in $C^{(2)}(T)$ can be formally
represented as  non-logarithmic renormalizations of $g_{1\parallel}$ and $%
g_{4\parallel}$. This renormalization involves the static bubble $\Pi
_{q=0}(q,\Omega =0)=T\sum_{\omega }\int dkG(k,\omega )G(k+q,\omega )$, which
is an anomaly by itself--it can be viewed as coming from low-energies (of
order $q$), if we sum over $\omega $ first, or from high energies, of order $%
\Lambda _{\mathrm{f}}$, if we integrate over $k$ first. It is then unclear
\emph{a priori }whether the scattering amplitudes in $C^{(2)}(T)$ are the
amplitudes on the scale of order $T$ or on the scale of the cutoff. To
verify this, we need to compute explicitly third-order diagrams.

\subsection{Third order diagrams and beyond}

\label{3rd}

\subsubsection{Third order diagrams}

\label{3rd_a}
\begin{figure}[tbp]
\begin{center}
\epsfxsize=1.0\columnwidth \epsffile{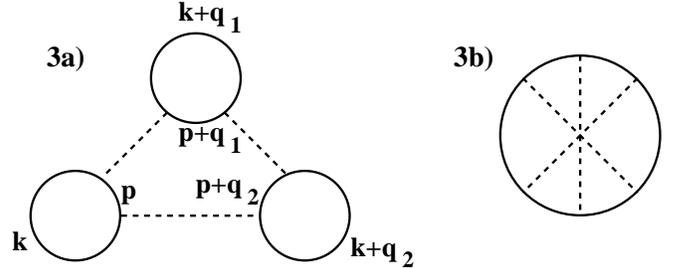}
\end{center}
\caption{Third-order diagrams that potentially give logarithmic
contributions. In diagram 3a),
 momenta $k$ and $p$ are counted from $\pm k_F
$, respectively. }
\label{fig:third}
\end{figure}
We analyze the third-order diagrams
in two steps.  At
the first step, we analyze possible logarithmic  terms in $\Xi $ at third
order,  searching for $\log T$ terms,  and also for terms which
contain $\log {\Lambda_{\mathrm{f}}/\Lambda_{\mathrm{b}}}$.
We will show that there are no $\log T$ terms at third order,
whereas all $\log {\Lambda_{\mathrm{f}}/\Lambda_{\mathrm{b}}}$ can be accounted
for by renormalizations of the $g_{1||,\perp}$ vertices.
At the second step, we will show that there exists a universal third-order term
which starts to flow at the next, fourth order.

 \paragraph{Logarithmic contributions.}
Diagrams that potentially contain logarithmic contributions are shown
 in Fig.~\ref
{fig:third}.
 Diagram 3a) is expressed via the cube of the polarization bubble at $2k_{F}$%
:
\begin{equation}
\Xi _{3a}^{(3)}=-\frac{\pi ^{2}}{3}~\left( g_{1\parallel
}^{3}+3g_{1\parallel }g_{1\perp }^{2}\right) T\sum_{\Omega }\int dq\Pi
_{2k_{F}}^{3}(q,\Omega ).  \label{44}
\end{equation}
The computation of $T\sum_{\Omega }\int dq\Pi _{2k_{F}}^{3}(q,\Omega )$ is
lengthy, and we present it in Appendix B. The result is
\begin{equation}
T\sum_{\Omega }\int dq\Pi _{2k_{F}}^{3}(q,\Omega )=\frac{T^{2}}{\pi }%
\left(L^2_{\mathrm{b}} - L_{\mathrm{b}}\right)  \label{45}
\end{equation}
where, we remind, $L_{\mathrm{b}} = \log \left(\Lambda_{\mathrm{f}}/\Lambda_{%
\mathrm{b}}\right)$. All potential $\log ^{2}T$ and $\log T$ terms cancel
out, and the only logarithmic dependence left involves the ratio of the
cutoffs. Substituting (\ref{45}) into (\ref{44}), we obtain
\begin{equation}
\Xi _{3a}^{(3)}=-\frac{\pi T^{2}}{3}~\left( g_{1\parallel
}^{3}+3g_{1\parallel }g_{1\perp }^{2}\right) \left( L^2_b - L_{\mathrm{b}}.
\right)  \label{46}
\end{equation}

Diagram 3b) is a vertex renormalization of the second-order diagram 2a) in
Fig.~\ref{fig:second}. The vertices in diagram 2a) can be both $g_1$ or one
of them can be $g_1$ and the other one $g_4$. The $2k_F$ bubble in diagram 3b)
of Fig.~\ref{fig:third} is inserted into the $g_1$ line in both cases. The
total result for diagram 3b) is
\begin{equation}
\Xi^{(3)}_{3b}=\frac{2\pi T^2}{3}\left(g_{1||}-g_{4||}\right)%
\left(g_{1||}^2+g_{1\perp}^2\right)L_{\mathrm{b}}.  \label{3_3b}
\end{equation}

Diagram 3c) in Fig.~\ref{fig:third} is a vertex renormalization of the
second-order diagram 2c) in Fig.~\ref{fig:second}. One of the lines of the
second-order diagram is $g_1$ and the other one is $g_2$. Inserting the $2k_F
$ bubble into the $g_1$ line, we obtain for diagram 3c)
\begin{equation}
\Xi^{(3)}_{3c}=-\frac{\pi T^2}{3}g_{2||}\left(g_{1||}^2+g_{1\perp}^2%
\right)L_{\mathrm{b}}  \label{3_3c}
\end{equation}

Note that only $2k_{F}$ couplings in $C(T)$ are renormalized. The $g_{2}$
coupling in $C(T)$ remains at its bare value. A renormalization of $g_{2}$
could potentially come from diagram 3d), but this diagram
 contains no $L_{\mathrm{b}}$ terms
 because all internal momenta in this diagram cannot deviate from external
momenta more than by $\Lambda _{\mathrm{b}}$, i.e., there is no space for
the logarithm in momentum integrals~\cite{comm_new}. Therefore, the logarithmic
part of  the third-order specific heat is obtained by combining the
results from Eqs.(\ref{46},\ref{3_3b}) and (\ref{3_3c})]
\begin{widetext}
\begin{equation}
C^{(3)}(T)=\frac{2\pi T}{3}~L^2_{\rm{b}}\left( g_{1\parallel
}^{3}+3g_{1\parallel }g_{1\perp }^{2}\right)-\frac{2\pi T}{3}~L_{\rm{b}}
\left[ g_{1\parallel
}^{3}+3g_{1\parallel }g_{1\perp }^{2}+2\left(g_{1||}-g_{4||}-\frac{1}{2}g_{2||}\right)\left(g_{1||}^2+g_{1\perp}^2\right)\right]
 \label{CT3}
\end{equation}
\end{widetext}
    The third-order specific heat can be obtained from the results
of the first and second orders by replacing the bare couplings $g_{1||,\perp }$
by their renormalized values, ${\tilde{g}}_{1||,\perp }$. In particular, the
$L_{\mathrm{b}}^{2}$ term in Eq.(\ref{CT3}) accounts for the third-order
ladder renormalization of $g_{1\parallel }$ ($g_{1\parallel }\rightarrow
(g_{1\parallel }^{3}+3g_{1\perp }^{2}g_{1\parallel })L_{b}^{2}$, see (\ref
{8a})] in the first order specific heat [Eq.~(\ref{24})]. The $L_{\mathrm{b}}
$ terms account for the renormalizations of the $g_{1||}$, $g_{1||}^{2}$,
and $g_{1\parallel }^{2}+g_{1\perp }^{2}$ terms in $C^{(2)}(T)$ [Eq.(\ref{24}%
)] according to
\begin{eqnarray*}
g_{1||} &\rightarrow &-\left( g_{1\parallel }^{2}+g_{1\perp }^{2}\right) L_{%
\mathrm{b}} \\
g_{1||}^{2} &\rightarrow &-2g_{1||}\left( g_{1\parallel }^{2}+g_{1\perp
}^{2}\right) L_{\mathrm{b}} \\
g_{1\parallel }^{2}+g_{1\perp }^{2} &\rightarrow &-2\left( g_{1\parallel
}^{3}+3g_{1\parallel }g_{1\perp }^{2}\right) L_{\mathrm{b}},
\end{eqnarray*}
[see Eqs.(\ref{8a},\ref{8b})].

\paragraph{Universal contributions.} To this end, we have not obtained a term with the running coupling on the
scale of $T$. We now demonstrate how such a term is generated at third
order. To do this,
we compute the constant, cutoff-independent term in $\Xi ^{(3)}$. We will
not attempt to calculate this term using Eq. (\ref{44}), as the calculations
are quite involved.  Rather, we assume, by analogy with the second-order
calculation, that this constant term is independent of the ratio of the
cutoffs and can be evaluated in the same computational procedure as the one
that led us to Eq.(\ref{43}), i.e.,  by reducing the $2k_{F}$ problem to the
small $q$ one, and representing the third-order  diagrams as the products of
two triads. The same procedure was employed by AE.

The relevant diagrams here are  diagrams 3a) and 3d). Using the
triad method, we obtain
 for their sum \begin{widetext}
\beq \Xi^{(3)}_{\mathrm{sum}} =\Xi _{3a}^{(3)}+\Xi _{3d}^{(3)}= -
\frac{4}{\pi} g_{1\parallel}g^2_{1\perp}
 ~T \sum_{\Omega_1} T \sum_{\Omega_2} \int dq_1 \int dq_2
\Pi_3 (q_1, q_2, \Omega_1, \Omega_2) \Pi_3 (-q_1, -q_2, \Omega_1,
\Omega_2), \label{48} \eeq where
 \bea &&\Pi_3 (q_1, q_2, \Omega_1, \Omega_2) =  T
\sum_{\omega_k}  \int dk
G_{R}(k, \omega_k)  G_{R}(k+ q_1,
\omega_k + \Omega_1)~  G_{R}(k+ q_2, \omega_k + \Omega_2), \nonumber \\
&&\Pi_3 (-q_1, -q_2, \Omega_1, \Omega_2) = T\sum_{\omega_p} \int
dip G_{L}(p, \omega_p)  G_{L}(p+ q_1, \omega_p + \Omega_1)~
G_{L}(p+ q_2, \omega_p + \Omega_2). \label{49} \eea
\end{widetext}
The integration in (\ref{49}) is straightforward, as all integrals converge,
and we have
\begin{eqnarray}
&&\Pi _{3}(q_{1},q_{2},\Omega _{1},\Omega _{2})=\frac{1}{2\pi }\left( \frac{%
i\Omega _{2}+q_{2}}{i\Omega _{2}-q_{2}}-\frac{i\Omega _{1}+q_{1}}{i\Omega
_{1}-q_{1}}\right)  \notag \\
&&\times ~\frac{1}{i(\Omega _{1}+\Omega _{2})-(q_{1}+g_{2})}.  \label{50}
\end{eqnarray}
Substituting into (\ref{48}), we obtain
\begin{widetext}
\bea \Xi^{(3)}_{\mathrm{sum}}   &=&  \frac{1}{\pi}
 g_{1\parallel}g^2_{1\perp} ~
\int dq_1 \int dq_2 T \sum_{\Omega_1} T \sum_{\Omega_2}
\left(\frac{\Omega_2}{i\Omega_2 -q_2} -
\frac{\Omega_1}{i\Omega_1 -q_1}\right)
\left(\frac{\Omega_2}{i\Omega_2 +q_2} -
\frac{\Omega_1}{i\Omega_1 + q_1}\right) \nonumber \\
&& \times
 ~\frac{1}{i(\Omega_1 + \Omega_2) - (q_1 + q_2)}  ~\frac{1}{i(\Omega_1 + \Omega_2) +  (q_1 +
 q_2)}.
\label{51}
\eea
\end{widetext}
The computation of the double momentum integral and frequency sum requires
special care. The most straightforward way is to sum over frequencies first,
as the frequency sums are not restricted by cutoffs. Performing the
summation, and using the symmetry between $q_{1}$ and $q_{2}$, we find after
some algebra
\begin{equation}
\Xi _{\mathrm{sum}}^{(3)}=\frac{1}{\pi }g_{1\parallel }g_{1\perp
}^{2}\int_{-\Lambda _{\mathrm{f}}}^{\Lambda _{\mathrm{f}}}dq_{1}\int_{-%
\Lambda _{\mathrm{f}}}^{\Lambda _{\mathrm{f}}}dq_{2}\left( \frac{1}{4}%
\right) .  \label{52}
\end{equation}
This obviously implies that the momentum integral is confined to high
energies, of order $\Lambda _{\mathrm{f}}$, and $\Xi _{a}^{(3)}$ does not
contain a $T^{2}$ term.

However, this is not the whole story. The new understanding is obtained if
we perform computations in different order, by integrating over momentum
first. The computation is again lengthy, but straightforward, and yields
\begin{eqnarray}
&&\Xi _{\mathrm{sum}}^{(3)}=\pi g_{1\parallel }g_{1\perp }^{2}T\sum_{\Omega
_{1}}T\sum_{\Omega _{2}}  \notag \\
&&\times \left( 1-\delta _{\Omega _{1},0}\delta _{\Omega _{2},0}\right)
F\left( \Omega _{1},\Omega _{2},\Lambda _{\mathrm{f}}\right),  \label{53}
\end{eqnarray}
where $\delta _{a,b}$ is the Kronecker symbol, and $F(\Omega _{1},\Omega
_{2},\Lambda _{\mathrm{f}})$ approaches a constant ($=1$) when frequencies
are much smaller than the fermionic cutoff $\Lambda _{\mathrm{f}}$. At
frequencies comparable and larger than the cutoff, $F$ is rather complex,
but the  part of $F$ relevant for our purposes is
\begin{equation}
F\left( \Omega _{1},\Omega _{2},\Lambda _{\mathrm{f}}\right) =1-\frac{3}{\pi
}\left[|\Omega _{1}|\frac{\Lambda _{\mathrm{f}}}{\Omega _{2}^{2}+\Lambda _{%
\mathrm{f}}^{2}} + |\Omega _{2}|\frac{\Lambda _{\mathrm{f}}}{\Omega
_{1}^{2}+\Lambda _{\mathrm{f}}^{2}}\right]  \label{54}
\end{equation}
There are other terms in $F$, but they do not lead to a $T^{2}$ term in $\Xi
_{\mathrm{sum}}^{(3)}$.

The double frequency sum in (\ref{53}) then reduces to
\begin{equation}
T\sum_{\Omega _{1}}T\sum_{\Omega _{2}}\left[ 1-\frac{6}{\pi }|\Omega _{1}|%
\frac{\Lambda _{\mathrm{f}}}{\Omega _{2}^{2}+\Lambda _{\mathrm{f}}^{2}}%
\right] -T^{2},  \label{55}
\end{equation}
where the summation is now over all Matsubara frequencies, including $\Omega
_{1},\Omega _{2}=0$  (we used the symmetry between $\Omega_1$ and $\Omega_2$%
).  The sum $T\sum_{\Omega _{1}}T\sum_{\Omega _{2}}1$ is confined to large
frequencies, and does not lead to $T^{2}$ term in $\Xi _{a}^{(3)}$. If $%
\Lambda _{\mathrm{f}}$ were infinite, $-T^{2}$ would be the only outcome of (%
\ref{55}). For finite $\Lambda _{\mathrm{f}}$, one has to be careful as the
second term in (\ref{55}) cannot be neglected for $\Omega _{2}\geq \Lambda _{%
\mathrm{f}}$. Replacing the sum over $\Omega _{2}$ by the integral, we
obtain that the contribution from the second term in (\ref{55}) reduces to $%
-(3/\pi )T\sum_{\Omega _{1}}|\Omega _{1}|=\mathrm{const}+T^{2}$.
Adding this result and the $-T^{2}$ term in (\ref{55}), we find
that the $T^{2}$ term in $\Xi _{a}^{(3)}$ vanishes. This agrees
with (\ref{52}). However, we now see that the vanishing of the
$T^{2}$ term in $\Xi _{a}^{(3)}$ is the result of a cancellation
between two physically different contributions. The $-T^{2} $ term
in (\ref{55}) is a truly low-energy contribution, which survives
even
if we set $\Lambda _{\mathrm{f}}=\infty $. This $T^{2}$ term comes from $%
\Omega _{1}=\Omega _{2}=0$, and from vanishingly small $q_{1}$, $q_{2}$ in
the momentum integrand. A very similar term leads to a non-analyticity in
the spin susceptibility ~\cite{cm}. The coupling for this term, $g_{1\perp
}^{2}g_{1\parallel }$, is then at the low-energy scale $(\sim T)$, and
should be fully renormalized within RG. On the other hand, the compensating $%
T^{2}$ term comes from large energies, of order $\Lambda _{\mathrm{f}}$, and
is therefore a high-energy contribution. The corresponding coupling is then
at the high energy scale, and it should remain constant under the RG
transformation.

As a result, $\Xi _{\mathrm{sum}}^{(3)}$ becomes
\begin{equation}
\Xi _{\mathrm{sum}}^{(3)}=-\pi T^{2}\left( g_{1\perp
}^{2}(L_{T})g_{1\parallel }(L_{T})-g_{1\perp }^{2}g_{1\parallel }\right) .
\label{56}
\end{equation}
where $g_{1}(L_{T})$ is the $2k_{F}$ coupling on the scale of $T$, and $%
g_{1} $ without argument is the coupling at the cutoff scale. The low-energy
contribution in (\ref{56}) coincides with the result obtained by AE (modulo
a factor of 2). AE did not evaluate the high-energy contribution in (\ref{56}%
). We did not attempt to obtain $\Xi _{a}^{(3)}$ for an arbitrary ratio of $%
\Lambda _{\mathrm{b}}$ and $\Lambda _{\mathrm{f}}$.  We expect that the
low-energy contribution is independent
of
the ratio of the cutoffs. At the same time, the high-energy term in (\ref{56}%
) may depend on the ratio of the cutoffs, i.e., the $(-1)$ factor between
low-energy and high-energy contributions in (\ref{56}) may only hold for $%
\Lambda _{\mathrm{b}}>\Lambda _{\mathrm{f}}$, when the ``triad'' calculation
is valid. In any event, the high-energy term in (\ref{56}) is a regular $%
T^{2}$ term and is therefore of little interest.

For completeness,  we
also note that there exists another high-energy contribution  of order $%
g_{1\perp }^{2} g_{1\parallel }$, obtained by inserting the first-order
renormalization of the Fermi velocity into the second-order backscattering
diagram. This contribution can be easily evaluated in the same way as (\ref
{56}) and yields
\begin{equation}
\Xi _{\mathrm{extra}}^{(3)}=-\pi T^{2}g_{1\perp }^{2}g_{1\parallel }
\label{57}
\end{equation}
If (\ref{56}) is independent on the ratio of the cutoffs, the
high-energy terms in (\ref{56}) and (\ref{57}) cancel each other,
i.e., the net result is only low-energy contribution. This
cancellation is likely accidental, however.

Assembling logarithmic and universal constant term at third order,
evaluating the specific heat, combining with first and second-order
diagrams, and using the RG flow of the couplings, we obtain the full result
for the specific heat $C(T)$, Eqs. (\ref{12}) and (\ref{14}).

\subsubsection{Fourth order diagrams}

\begin{figure}[tbp]
\begin{center}
\epsfxsize=0.5\columnwidth \epsffile{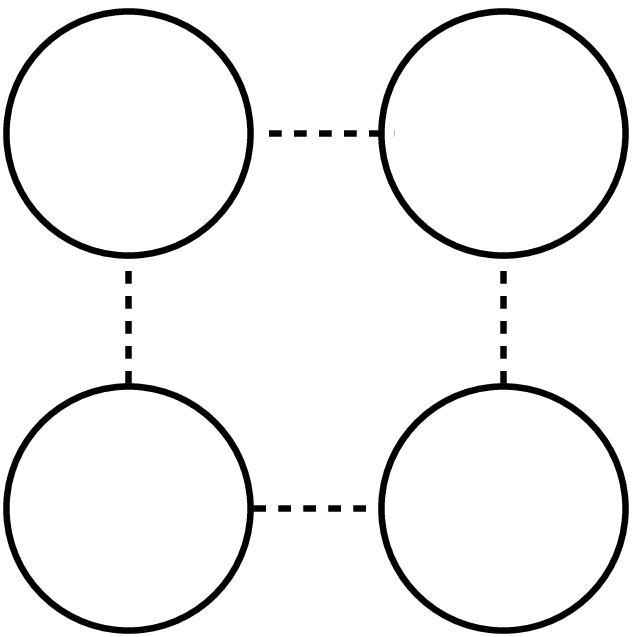}
\end{center}
\caption{Fourth-order diagram with four bubbles.}
\label{fig:fourth}
\end{figure}
For completeness, we also computed explicitly the fourth-order, four-bubble
backscattering diagram for the thermodynamic potential  We indeed found a $%
T^2 \log T$  term obtained by combining the "zero-energy" contribution to $%
\Xi_{\mathrm{sum}}^{(3)}$ [Eq.~(\ref{56})] with  an additional polarization
bubble $\Pi _{2k_{F}}(0,0)\propto \log T$. This one accounts for the $\log T$
renormalization of the running couplings $g_{1\perp }(L_{T})$ and $%
g_{1\parallel }(L_{T})$ in (\ref{56}). We searched for possible other $T^2
\log T$ contributions,  using the same method as at the end of the previous
section. Namely, we assumed that $T^2 \log T$ terms must be independent of
the cutoff ratio, set $\Lambda _{\mathrm{b}}>\Lambda _{\mathrm{f}}$,  and
created two ``quaternions'' by assembling four fermionic propagators with
close momenta $k\approx k_{F}$, $k+q_{1}$, $k+q_{2}$, $k+q_{3}$, and $%
p\approx -k_{F}$, $p+q_{1}$, $p+q_{2}$, $p+q_{3}$, $|q|_{i}\ll k_F$ [see Fig.%
\ref{fig:fourth}]. We integrated independently over $k$ and $p$ in infinite
limits (we recall that this is possible only if the cutoff imposed by the
interaction is irrelevant), then integrated over $q_{i}$ and summed over
corresponding frequencies. We found $T^2 \log T$ terms from particular
regions of frequency summations and integrations over three bosonic momenta $%
q_{1}$, $q_{2}$, and $q_{3}$; however,  all such $T^2 \log T$ terms cancel
out.  Therefore, the only non-vanishing $T^2\log T$ contribution at fourth
order is the "zero-energy" one.  All $T^{2}$ terms in $\Xi $ up to fourth
order are anomalies, and corresponding couplings are at energies $O(\Lambda )
$.

\section{Comparison to the sine-Gordon model}

\label{sec:sine}

\subsection{Model}

A well-established way to treat the system of 1D fermions is bosonization,
which allows one to map the original problem onto the quantum sine-Gordon
model. It is instructive to see how the results of the previous Sections can
be obtained within this model.  We bosonize the operators of right- and
left-moving fermions, $R_{\alpha }$ and $L_{\alpha }$, in a standard way
\begin{equation*}
R_{\alpha }(x),L_{\alpha }(x)=\frac{1}{\sqrt{2\pi a}} \exp \left[ \pm
i\left( \phi _{\alpha }(x)\mp \theta _{\alpha }(x)\right) \right] ,\text{ }%
\alpha =\uparrow ,\downarrow ,
\end{equation*}
where $a$ is a short-distance cutoff related to the momentum cutoff
introduced in the previous Sections via $a=\Lambda _{\text{f}}^{-1}.$ Upon
bosonization, the part of the fermionic Hamiltonian  parameterized by
couplings $g_{4}$ and $g_{2}$ is mapped onto the Gaussian part of the
bosonic Hamiltonian, $H_{G}=H_{G}^{\left( \rho \right) }+H_{G}^{\left(
\sigma \right)}$, where
\begin{widetext}
\begin{eqnarray}
H_{\mathrm{G}}^{\left( \rho ,\sigma \right) } &=&\frac{1}{2}\int
dx\left( 1+g_{4||}\pm g_{4\perp }+g_{2||}\pm g_{2\perp }\right)
\left( \partial _{x}\phi _{\rho ,\sigma }\right) ^{2}+\left(
1+g_{4||}\pm g_{4\perp }-g_{2||}\mp g_{2\perp }\right) \left(
\partial _{x}\theta _{\rho ,\sigma }\right) ^{2},\label{gauss}
\end{eqnarray}
\end{widetext}
and the charge and spin bosons are defined as $\phi _{\rho ,\sigma }=\left(
\phi _{\uparrow }\pm \phi _{\downarrow }\right) /\sqrt{2}$ and $\theta
_{\rho ,\sigma }=\left( \theta _{\uparrow }\pm \theta _{\downarrow }\right) /%
\sqrt{2}$. The $2k_{F}$ scattering, however, leads to non-linear, cosine
terms in the bosonic Hamiltonian. For a local, delta-function interaction,
the cosine term only  comes from $2k_{F}$-scattering of fermions with
opposite spins (coupling $g_{1\perp }$). However, for an arbitrary,
non-local interaction, there is also a cosine term which  comes from $2k_{F}$%
-scattering of fermions with parallel spins (coupling $g_{1\parallel }$).
Introducing a finite-range interaction $V_{12}\equiv V(x_{1}-x_{2})$, we map
the $2k_{F}$ part of the fermionic Hamiltonian onto
\begin{widetext}
\begin{eqnarray}
 H_{1||,\perp}=\frac{2}{\left( 2\pi a \right) ^{2}}\int\int
dx_{1}dx_{2}V_{12}\cos \left[ \sqrt{2\pi }\left( \phi _{\rho
}(x_{1})-\phi _{\rho }\left( x_{2}\right) \right) +2k_{F}\left(
x_{1}-x_{2}\right) \right]\cos \left[ \sqrt{2\pi }\left( \phi
_{\sigma }(x_{1})\mp\phi _{\sigma }\left( x_{2}\right) \right)
\right]. \label{sG}
\end{eqnarray}
\end{widetext}
For a local interaction, $V_{12}=V_{0}\delta (x_{1}-x_{2})$, Eq. (\ref{sG})
reduces to the usual sine-Gordon model.

The universal $g_{1}^{3}$ term in the thermodynamic potential (the analog of
the universal term in Eq. (\ref{56}) for the SU(2) symmetric case) was
obtained by Cardy \cite{cardy} and Ludwing and Cardy~\cite{ludwig_cardy}
for a general case of a conformal theory perturbed about a fixed point by a
marginally irrelevant operator, and we just refer the reader to that work.
The first and second-order terms in $g_{1}$, however, have not been obtained
explicitly in the sine-Gordon model before.  Our goal is to demonstrate how
the anomalous terms of order $g_{1}$ and $g_{1}^{2}$ appear in the
thermodynamic potential, and, in particular, how the $g_{1}$ coupling gets
a logarithmic renormalization on a scale of the bosonic cutoff in this
model. We will see that to get this renormalization, and also to obtain $%
g_{1}^{2}$ term with a correct prefactor, one \textit{must} consider a
finite-range interaction and keep the range of the interaction larger than
the short-distance cutoff of the theory.

The thermodynamic potential per unit length is given by
\begin{equation}
\Xi=-\frac{T}{L}\log \int D\phi \exp \left( -\left[ S_{G}+S_{1||}+S_{1\perp }%
\right] \right) ,  \label{xi_bos}
\end{equation}
where $S_{a}$, with $a={\mathrm{G}},1||$, and $1\perp $, are the
actions corresponding to the Gaussian and $2k_{F}$ parts of the
bosonic Hamiltonian, respectively. Expansion in $S_{1||}+S_{1\perp
}$ generates perturbation series for $\Xi $.  In the absence of
backscattering ($g_{1}=0$), the bosons are free and theory is
exactly solvable for arbitrary $g_{2}$ and $g_{4}$. One can then
construct the perturbation theory in $ g_{1||}$ and $g_{1\perp } $
about the free-boson point.  To make a connection with the
previous Sections, however, we will perform the perturbative
expansion in all coupling constants rather than only in $g_1$.
This means that the averages generated by an expansion in
$S_{1||}$ and $S_{1\perp }$ will be taken over a free Gaussian
action,  Eq. (\ref{gauss}) with $g_4 = g_2 =0$.

\subsection{First order}

\label{sec:first} The first-order term is obtained by expanding the
exponential in (\ref{xi_bos}) to first order in $S_{1||}$. Performing the
averaging, we obtain
\begin{equation}
{\Xi}^{(1||)}=2\int_{|x|\geq a}dxV\left( x\right) A_{\rho }(x,0)A_{\sigma
}(x,0)\cos (2k_{F}x),  \label{xi_1}
\end{equation}
where
\begin{equation}
A_{\rho ,\sigma }(x,\tau )=\frac{1}{\pi a}\langle e^{i\sqrt{2\pi }\left[
\phi _{\rho ,\sigma }(x,\tau )-\phi _{\rho ,\sigma }(0,0)\right] }\rangle .
\end{equation}
As the averages are calculated over a free Gaussian action, $A_{\rho }$ and $%
A_{\sigma }$ are equal to each other and given by \cite{giamarchi_book}
\begin{equation}
A_{\rho }(x,\tau )=A_{\sigma }(x,\tau )=\left[ \frac{1}{4}\frac{T^{2}}{\sinh
^{2}\pi xT+\sin ^{2}\pi \tau T}\right] ^{1/2}.
\end{equation}
The $2k_{F}$ polarization bubble in the $x,\tau $ space is
\begin{equation}
\Pi _{2k_{F}}\left( x,\tau \right) =-A^2_{\rho ,\sigma }\left( x,\tau
\right) =-\frac{1}{4}\frac{T^{2}}{\sinh ^{2}\pi xT+\sin ^{2}\pi \tau T}.
\label{aa_2}
\end{equation}
Therefore, the first-order result reduces to
\begin{equation}
{\Xi}^{(1||)}=-2\int dxV\left( x\right) \Pi _{2k_{F}}\left( x,0\right) \cos
(2k_{F}x).
\end{equation}
Note that this is nothing more than the first-order diagram 1a) in Fig. \ref
{fig:first}, written in the $x,\tau $ space. Expanding $\Pi _{2k_{F}}\left(
x,0\right) $ for $x\ll T^{-1},$ we obtain
\begin{equation}
\Pi _{2k_{F}}\left( x,0\right) =-\frac{T^{2}}{4\sinh ^{2}\pi xT}=-\frac{1}{%
4\pi ^{2}x^{2}}+\frac{1}{12}+\dots  \label{pi_x}
\end{equation}
The universal, constant term in $\Pi _{2k_{F}}\left( x,0\right) $ gives a $T$
dependent part of ${\Xi}^{(1)}_{2k_{F}}$%
\begin{equation}
{\Xi}^{(1||)}=\mathrm{const-}\frac{\pi }{3}g_{1||}T^{2},  \label{xi_1_bos}
\end{equation}
where $g_{1||}=(1/2\pi) \int dxV\left( x\right) \cos [2k_{F}x]$. We see that
the bosonization result, Eq.~(\ref{xi_1_bos}), agrees with the diagrammatic
one, Eq.~( \ref{23_3}), obtained for $\Lambda _{\mathrm{b}}<\Lambda _{%
\mathrm{f}}.$  This last condition is implicit in bosonization, as Eq. (\ref
{aa_2})  is valid only if the fermionic cutoff exceeds the bosonic one.
Notice that the final result Eq.~(\ref{xi_1_bos})  is formally valid also
for a local interaction.  However, the limit of a local interaction cannot
be taken at the very beginning. Indeed, in this limit $S_{1||}$ reduces to a
constant and does not contribute to the $T$ dependence of $\Xi $.

\subsubsection{Gaussian form of $H_{1||}$}

The $g_{1||}$ term in the specific heat can be also obtained by reducing $%
H_{1||}$ in Eq.~(\ref{sG}) to the Gaussian form, similar to what was done in
Ref.~[\onlinecite{capponi}] for spinless fermions.  For completeness, we
repeat this derivation here for fermions with spin. Indeed, $H_{1||}$ can be
written as the convolution of the $2k_{F}$ components of the density
\begin{equation}
H_{1||}=\frac{1}{2}\sum_{\alpha =\uparrow ,\downarrow }\int dx_{1}\int
dx_{2}V_{12}\rho _{2k_{F},\alpha }(x_{1})\rho _{2k_{F},\alpha }\left(
x_{2}\right) ,
\end{equation}
where $\rho _{2k_{F},\alpha }(x)=R_{\alpha }\left( x\right) L_{\alpha
}^{\dagger }\left( x\right) e^{2ik_{F}x}+\text{H.c}=\left(e^{2i\sqrt{\pi }%
\phi _{\alpha }\left( x\right) }e^{2ik_{F}x}+\text{H.c.}\right)/2\pi a.$
Performing normal ordering in the product of two exponentials of the bosonic
field and Taylor expanding the difference $\phi _{\alpha }\left(
x_{1}\right) -\phi _{\alpha }\left( x_{2}\right) $ under the normal-ordering
sign, one obtains
\begin{widetext}
\begin{eqnarray*}
e^{2i\sqrt{\pi }\phi _{\alpha }\left( x_{1}\right) }e^{-2i\sqrt{\pi }\phi
_{\alpha }\left( x_{1}\right) } =:e^{2i\sqrt{\pi }\left[ \phi _{\alpha
}\left( x_{1}\right) -\phi _{\alpha }\left( x_{2}\right) \right] }:
\exp %
\left[ ^{-4\pi \langle \phi \left( _{\alpha }\left( x_{1}\right) -\phi
_{\alpha }\left( x_{2}\right) \right) ^{2}\rangle }\right]  \\
=:1-2\pi \left( \partial _{x}\phi _{\alpha }\right) ^{2}\left(
x_{1}-x_{2}\right) ^{2}+\dots :\frac{a^{2}}{\left(
x_{1}-x_{2}\right) ^{2}} =c\text{-number}-2\pi a^{2}\left( \partial
_{x}\phi _{\alpha }\right) ^{2}.
\end{eqnarray*}
\end{widetext}
At the level of operators, $H_{1||}$ then reduces to
\begin{equation*}
H_{1||}=-g_{1||}\sum_{\alpha }\int dx\left( \partial _{x}\phi _{\alpha
}\right) ^{2}.
\end{equation*}
Combining this result with the Gaussian part of the Hamiltonian, we obtain a
new effective Hamiltonian

\begin{equation*}
H_{G}^{\ast }=\frac{1}{2}\sum_{\nu =\rho ,\sigma }\int_{x}\left( u_{\nu
}/K_{\nu }\right) \left( \partial _{x}\phi _{\nu }\right) ^{2}+\left( u_{\nu
}K_{\nu }\right) \left( \partial _{x}\theta _{\nu }\right) ^{2},
\end{equation*}
where the charge and spin velocities are the same as in Eq.(\ref{15}) and

\begin{equation}
K_{\rho ,\sigma }=\left( \frac{1+g_{4||}\pm g_{4\perp }-g_{2||}\mp g_{2\perp
}}{1+g_{4||}\pm g_{4\perp }+g_{2||}\pm g_{2\perp }-2g_{1||}}\right) ^{1/2}.
\label{k_1}
\end{equation}
The specific heat, corresponding to $H_{G}^{\ast },$ is given by the first
term in Eq.(\ref{16}). Notice that, in contrast to conventional bosonization
which treats the $g_{1||}$ interaction only as an exchange process to the $%
g_{2||}$ interaction and, therefore, contains only a combination of $%
g_{2||}-g_{1||},$ $K_{\rho,\sigma}$ in (\ref{k_1}) contain two combinations:
$g_{2||}-g_{1||}$ and $g_{4||}-g_{1||}.$

\subsection{Second order}

\label{sec:second}

We next demonstrate how the renormalization of the $g_{1\parallel}$ occurs
in  the bosonic language, and how the ``universal" term in $\Xi$ with $%
g^2_{1\perp}$ emerges within the sine-Gordon model. Expanding Eq.(\ref
{xi_bos}) to order $S_{1\perp }^{2}$ and performing the averaging, we obtain
the second-order piece in $\Xi $%
\begin{widetext}
\begin{equation} \Xi^{\left( 2 \right) }=-\int dx_1\int dx_2\int dx_3V_{13}V_{23}\cos
[2k_{F}\left( x_1-x_2\right]J_{\tau}(x_1,x_2), \label{xi_bos_2}
\end{equation}
\end{widetext}
where
\begin{equation}
J_{\tau }(x,x^{\prime })=\int_{0}^{1/T}d\tau \Pi _{2k_{F}}\left( x,\tau
\right) \Pi _{2k_{F}}\left( x^{\prime },\tau \right),  \label{int_tau}
\end{equation}
and all spatial integrals are cut at small distances by $a.$ Again, this is
nothing more that the two-bubble diagram 2b) in Fig.\ref{fig:second},
written in the $x,\tau $ space. The integration over $\tau $ is readily
performed
\begin{equation*}
J_{\tau }\left( x,x^{\prime }\right) =\frac{T^{3}}{4}\frac{\coth \left[ \pi
T\left( \left| x\right| +\left| x^{\prime }\right| \right) \right] }{\sinh
\left( 2\pi T\left| x\right| \right) \sinh \left( 2\pi T\left| x^{\prime
}\right| \right) }.
\end{equation*}
There are two contributions to the $T^{2}$ term in $\Xi^{\left( 2\right) }:$
one comes from large distances $|x|\sim |x^{\prime }|\sim T^{-1}$ and
another one comes from distances of the order the interaction range. For the
first contribution, the requirement that the potential must have a finite
range is irrelevant, and the  interaction in (\ref{xi_bos_2}) can be safely
replaced by the delta-function $V\left( x\right) =2\pi g_{1\perp }\delta
\left( x\right)$. We then obtain
\begin{eqnarray}
\Xi_{a}^{\left( 2\right) } &=&-2\pi ^{2}g_{1\perp }^{2}T^{3}\int_{a}^{\infty
}dx\frac{\coth \left( 2\pi Tx\right) }{\sinh ^{2}\left( 2\pi Tx\right) } ,
\notag \\
&=&-\frac{\pi }{2}g_{1\perp }^{2}\frac{T^{2}}{\sinh ^{2}\left( 2\pi
Ta\right) }.
\end{eqnarray}
Expanding the last result for $Ta\ll 1$, we obtain
\begin{equation}
\Xi_{a}^{(2)}=\mathrm{const}+\frac{\pi }{6}g_{1\perp }^{2}T^{2}.
\label{xi_univ_1}
\end{equation}
This contribution is of the same magnitude but opposite in sign to the
cutoff-independent part of $\Xi $ in Eq.(\ref{41}). As the second
contribution is expected to come from distances smaller than $T^{-1},$ we
expand Eq.(\ref{int_tau}) for $T\rightarrow 0$ and keep only the $T$
dependent term
\begin{equation*}
J_{\tau }\left( x,x^{\prime }\right) =-\frac{T^{2}}{48\pi }\frac{\left(
\left| x\right| -\left| x^{\prime }\right| \right) ^{2}}{\left| x\right|
\left| x^{\prime }\right| \left( \left| x\right| +\left| x^{\prime }\right|
\right) }.
\end{equation*}
Introducing new variables $\xi =x-x^{\prime }$, $\eta =\left( x+x^{\prime
}\right) /2$ and performing elementary integrations, the resulting
contribution to $\Xi $ can be represented as a sum of two terms
\begin{eqnarray}
\Xi_{b}^{\left( 2\right) } &=&\Xi _{+}+\Xi _{-},  \notag \\
\Xi _{+} &=&\frac{T^{2}}{12\pi }\int_{2a}^{\infty }d\eta W\left( \eta
\right) \cos \left( 2k_{F}\eta \right) F_{+}\left( \eta \right),  \notag \\
\Xi _{-} &=&\frac{T^{2}}{12\pi }\int_{0}^{\infty }d\xi W\left( \xi \right)
\cos \left( 2k_{F}\xi \right) F_{-}\left( \xi \right),
\end{eqnarray}
where
\begin{equation}
F_{+}\left( \eta \right) =\log \frac{\eta -a}{a}-2+4\frac{\eta }{a},\text{ }%
F_{-}\left( \xi \right) =\log \frac{\left( \xi /2+a\right) }{a\left( \xi
+a\right) }  \label{fpm}
\end{equation}
and $W(x)=\int dy V(x+y)V(y)$. The universal, cutoff-independent part of $%
\Xi_{b}^{\left( 2\perp \right) }$ comes from the constant term ($-2$) in the
function $F_{+}\left( \eta \right) .$ It is of the opposite sign and twice
larger than the contribution in Eq.(\ref{xi_univ_1}). Combining these two
contributions together, we obtain for the universal part of $\Xi $%
\begin{equation*}
\Xi_{\mathrm{univ}}^{\left( 2\right) }=-\frac{\pi }{6}g_{1\perp }^{2}T^{2}.
\end{equation*}

The remainder of $\Xi $ is a cutoff-dependent part. To calculate this part,
we consider two model interactions. The first one is consistent with the
assumption used in the previous Sections (and also in $g$ -ology, in
general) that the backscattering part of the interaction is peaked near $%
2k_{F},$ i.e., the interaction oscillates in real space with period $\pi
/k_{F}.$ A model which describes this behavior is
\begin{equation*}
V\left( x\right) =g_{1\perp }\frac{2b}{x^{2}+b^{2}}\cos \left(
2k_{F}x\right) .
\end{equation*}

The scale $b$ equals to the bosonic cutoff $\Lambda^{-1}_{\mathrm{b}}$
introduced earlier. The assumption $\Lambda _{\mathrm{b}}\ll \Lambda _{%
\mathrm{f}}$ corresponds to the condition $b\gg a.$ Expanding functions $%
F_{\pm }$ for $\xi ,\eta \gg a$ and neglecting the exponentially small terms
(of order $\exp \left( -2k_{F}b\right) )$ as well as terms proportional to
powers of $a,$ we arrive at
\begin{equation*}
\Xi_{\mathrm{nonuniv}}^{\left( 2\right) }=\frac{4T^{2}}{3}g_{1\perp
}^{2}b\int_{0}^{\infty }\frac{d\xi }{\xi ^{2}+\left( 2b\right) ^{2}}\log
\frac{\xi }{2a}=\frac{\pi T^{2}}{3}g_{1\perp }^{2}\log \frac{b}{a},
\end{equation*}
where we used that $\int_{0}^{\infty }dx\log x/\left( 1+x^{2}\right) =0.$
Combining the universal and non-universal parts together, we obtain

\begin{equation}
\Xi^{\left( 2\right) }=-\frac{\pi }{3}g_{1\perp }^{2}T^{2}\left( 1-2\log
\frac{b}{a}\right) ,  \label{41_bos}
\end{equation}
which coincides with Eq.(\ref{41})
upon identifying $\log \frac{b}{a}=L_{\mathrm{b}}$.

We see that the logarithmic renormalization of the backscattering coupling
is reproduced within the sine-Gordon model. However, this result could  have
not be obtained for a  local  interaction. The interaction must have a
finite range,  which is larger than the short-distance cutoff in the theory.

Another model, which we consider for completeness, corresponds to a
long-range potential, i.e., to an interaction peaked near $q=0$ in the
momentum space. To describe this situation, we choose
\begin{equation*}
V\left( x\right) =\frac{u}{\pi }\frac{b}{x^{2}+b^{2}},
\end{equation*}
and assume that $b\gg a\sim k_{F}^{-1},$ so that the $2k_{F}$ component of
the potential $V(2k_F)=u\exp(-2k_Fb)$ is exponentially small. Such
interaction is not considered in the $g$ -ology, and we will not express its
parameters in terms of $g$ -couplings. If backscattering is neglected
completely, the problem is exactly soluble either via bosonization or
Dzyaloshinskii-Larkin diagrammatic formalism \cite{DL}. It turns out that,
somewhat surprisingly, corrections to the exact solution are small not
exponentially but only algebraically, in parameter $1/k_{F}b.$ The reason is
that logarithmic terms in functions $F_{\pm }$ [cf.Eq.(\ref{fpm})], which
reflect correlations in motion of free fermions, introduce branch cuts into
the integrals. The contribution of these branch cuts to the result is much
larger than the exponentially small contribution of the poles in the
interaction potential. Evaluating the integrals and keeping only the leading
terms, we arrive at
\begin{equation*}
\Xi^{\left( 2\right) }=\frac{T^{2}}{48\pi ^{2}}u^{2}\left[ \frac{\sin 4k_{F}a%
}{bk_{F}}\ln \frac{2eb}{a}-\frac{\cos 4k_{F}a}{2k_{F}^{2}ab}+O\left( \frac{1%
}{k_{F}^{4}a^{2}b^{2}}\right) \right] ,
\end{equation*}
where $e=2.718..$. The universal term, which is proportional to the $2k_{F}$
component of the interaction is exponentially small and we do not retain it
here.

In the opposite case of a short-range interaction, i.e., for $b\ll a\sim
k_{F}^{-1},$ $\Xi^{\left( 2\right) }$ is given entirely by the universal
term
\begin{equation*}
\Xi^{\left( 2\right) }=-\frac{u^{2}}{12\pi }T^2.
\end{equation*}

\section{Conclusions}

\label{conclusions}  In conclusion, we  performed a detailed
analysis of the temperature dependence of the specific heat for a
1D interacting Fermi system.  We used the $g-$ology model, and
carried out a perturbative expansion in the  couplings in the
fermionic language. We have shown that, to first two orders in the
interactions,  the specific heat is expressed in terms of the
non-running couplings in the RG -sense. The $g_4 $, and the $g_2$
vertices appearing in $C(T)$ are just bare  vertices, while the
backscattering $g_1$ vertex is the effective one,  renormalized by
fermions with momenta between fermionic and bosonic cutoffs. The
running backscattering amplitude on the scale of $T$  appears in
the specific heat only at third order in perturbation theory,  The
$\log T$ renormalization of the specific heat at the lowest $T$,
expected from the RG flow of the coupling constants, then only
occurs at the fourth order in the perturbation theory, and the
$T$-dependence of the specific heat follows the RG flow of the
cube of the backscattering amplitude, in agreement with previous
studies.  We explicitly demonstrated that the absence of the
logarithmic corrections below fourth order is due to cancellation
of $\log T$ terms coming from low energies, of order $T$,  and
high energies, of order of the ultraviolet cutoffs in the theory.
We also showed how the diagrammatic results can be obtained within
the sine-Gordon model.

\acknowledgments
We acknowledge helpful discussions with I. L. Aleiner,
C. Castellani,  F. Essler, A. M. Finkelstein, T. Giamarchi, L. I. Glazman,
K. B. Efetov, A. W. W. Ludwig, K. A. Matveev, A. A. Nersesyan, G. Schwiete,
O. A. Starykh, and G. E. Volovik, support from NSF-DMR 0604406 (A. V. Ch.),
0308377 (D. L. M.), 0529966 and 0530314 (R.S.), and the hospitality of the
Aspen Center of Physics. D. L. M. and R. S. acknowledge the hospitality of
the ICTP (Trieste, Italy), where part of their work was done. A.V. Ch
acknowledges the hospitality of the TU Braunschweig during the completion of
this work.

\section{Appendix A}

\label{appA}

\begin{figure}[tbp]
\begin{center}
\epsfxsize=0.5\columnwidth \epsffile{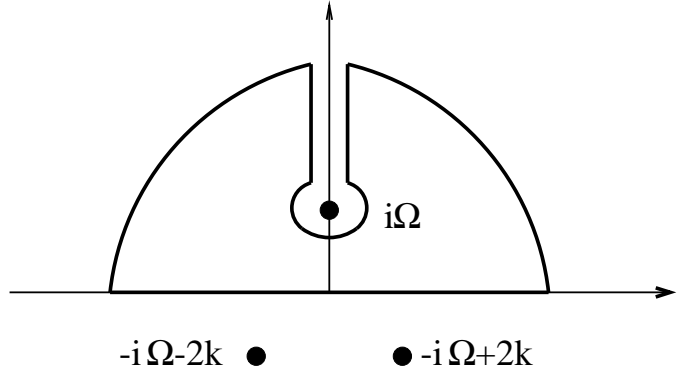}
\end{center}
\caption{Integration contour for Eq.(\ref{a5}). }
\label{fig:contour}
\end{figure}

In this Appendix, we derive Eq. (\ref{40}) for $X\equiv T\sum_{\Omega }\int
dq\Pi _{2k_{F}}^{2}(q,\Omega )$. The polarization operator $\Pi
_{2k_{F}}(q,\Omega )$ is given by (\ref{26}). It is convenient to split $X$
into three terms $X=X_{1}+X_{2}+X_{3}$ as
\begin{widetext}
\bea
&& X_1 = T \sum_\Omega \int_0^{\Lambda_{\mathrm{b}}} \frac{dq}{2\pi^2} \log^2 {\frac{\Omega^2 + q^2}{4\Lambda^2_f}}, \nonumber \\
&& X_2 = - \frac{8}{\pi^2}  T \sum_\Omega \int_0^{\Lambda_{\mathrm{b}}} dq
\log{\frac{\Omega^2 +q^2}{4 \Lambda^2_f}} \int_0^\infty dx
 dk k n_F (k)  \left(\frac{1}{(q-i\Omega)^2 - 4 k^2}
  + \frac{1}{(q + i \Omega)^2 - 4 k^2}\right), \nonumber \\
&& X_3 = \frac{32}{\pi^2}  T \sum_\Omega \int_0^{\Lambda_{\mathrm{b}}} dq
\left[\int_0^\infty dk k n_F (k) \left(\frac{1}{(q-i\Omega)^2 - 4
k^2}
 + \frac{1}{(q + i \Omega)^2 - 4 k^2}\right)\right]^2.
\label{a1} \eea
\end{widetext}
As we are only interested in a finite $T$ contribution, we can safely
subtract $T\sum_{\Omega }\int_{0}^{\infty }\frac{dq}{2\pi ^{2}}\log ^{2}{%
\frac{q^{2}}{4\Lambda _{\mathrm{f}}^{2}}}$ from $X_{1}$. The rest is
ultraviolet-convergent, and we can integrate explicitly over $q$ by setting
the upper limit of the $q-$integral to infinity. We obtain
\begin{equation}
X_{1}=\frac{2}{\pi }T\sum_{\Omega }|\Omega |\left[ \log {\frac{|\Omega |}{%
2\Lambda _{\mathrm{f}}}}-1+\log {2}\right] .  \label{a2}
\end{equation}
Using
\begin{eqnarray}
&&T\sum_{\Omega }|\Omega |=-\frac{\pi T^{2}}{3} ,  \notag \\
&&T\sum_{\Omega }|\Omega |\log {|\Omega |}=-\frac{\pi T^{2}}{3}\log T+\frac{%
\pi T^{2}}{6}-\frac{2T^{2}}{\pi }I_{1} ,  \notag \\
&&I_{1}=\int_{0}^{\infty }\frac{x^{2}\log {2x}}{\sinh ^{2}x}=0.5803,
\label{a3}
\end{eqnarray}
we obtain
\begin{equation}
X_{1}=\frac{2T^{2}}{3}\left[ -\log {\frac{T}{2\Lambda _{\mathrm{f}}}}+ B%
\right],  \label{a4}
\end{equation}
where $B = \frac{3}{2}-\log 2-\frac{6}{\pi ^{2}}I_{1} = 0.454$. The second
term, $X_{2}$, contains contributions both from small $q$, of order $T$, and
from large $q$, of order $\Lambda _{\mathrm{b}}$. It is convenient to split
the momentum integral $\int_{0}^{\Lambda _{\mathrm{b}}}$ into $%
\int_{0}^{\infty }-\int_{\Lambda _{\mathrm{b}}}^{\infty }$. The first
integral can be easily converted into the integral over the whole real $q$
axis. The poles in $q$ at any finite $\Omega $ are located in the same
half-plane, and the $q-$integral is nonzero only because of the branch cut
in the logarithm. Choosing the integration contour as shown in Fig.~\ref
{fig:contour}, evaluating the momentum integral, performing the frequency
sum, and adding a separate contribution from $\Omega =0$, we obtain
\begin{widetext}

\beq - \frac{8}{\pi^2}  T \sum_\Omega \int_0^\infty dq
\log{\frac{\Omega^2 +q^2}{4 \Lambda^2_f}} \int_0^\infty dk k
 n_F (k)\left(\frac{1}{(q-i\Omega)^2 - 4 k^2} + \frac{1}{(q + i \Omega)^2 - 4 k^2}\right) =
  \frac{2 T^2}{3} \left[\log{\frac{T}{2\Lambda_{\mathrm{f}}}}
 -B -(0.5+\log 2) \right]. \label{a5} \eeq
\end{widetext}
The integral over large $q>\Lambda _{\mathrm{b}}$ involves also large
frequencies $\Omega \sim q$, and the frequency sum can be safely replaced by
the integral. Typical fermionic momenta, $k$, are of order $T$ and,
therefore, much smaller than $q$ and $\Omega $. Neglecting $k$ in the
denominators of the integrand, and performing three independent integrations
(over $k$, $\Omega $, and $q$), we obtain
\begin{widetext}
\bea && \frac{8}{\pi^2}  T \sum_\Omega \int_{\Lambda_{\mathrm{b}}}^\infty dq
\log{\frac{\Omega^2 +q^2}{4 \Lambda^2_f}} \int_0^\infty dk k
 n_F (k) \left(\frac{1}{(q-i\Omega)^2 - 4 k^2} + \frac{1}{(q + i \Omega)^2 - 4 k^2}\right) \nonumber \\
&&=  \frac{8}{\pi^3} \int_0^\infty dk k
 n_F (k)  \int d \Omega  \int_{\Lambda_{\mathrm{b}}}^\infty dq \frac{1}{(q-i\Omega)^2}  = - \frac{2T^2}{3} \left[\log{\frac{2\Lambda_{\mathrm{f}}}{\Lambda_{\mathrm{b}}}} -
 -(\frac{1}{2} + \log{2}) \right]. \label{a6} \eea
\end{widetext}
Combining the two contributions, we obtain
\begin{equation}
X_{2}=\frac{2T^{2}}{3}\left[ \log {\frac{T}{2\Lambda _{\mathrm{f}}}}-\log {%
\frac{2\Lambda _{\mathrm{f}}}{\Lambda _{\mathrm{b}}}}-B\right] .  \label{a7}
\end{equation}
In the third term, $X_{3}$, the 2D integral over $q$ and $\Omega $ is
ultraviolet convergent, and we can safely set the upper limit of the $q-$%
integral to infinity. The integral again has separate contributions from $%
\Omega \neq 0$, and from $\Omega =0$. The contribution to $X_3$ from finite
frequencies is evaluated straightforwardly by closing the contour of the $q-$%
integral in the upper or lower half-plane. We obtain $(T^2/3) (7-10 \log 2)$%
. The evaluation of the contribution from $\Omega =0$ requires special care
because of the poles which are avoided by replacing $\Omega$ by $i\delta$.
The corresponding contribution to $\Xi_3$ becomes
\begin{widetext}
\bea && \frac{32}{\pi^2}~ T \int_0^\infty dq \int dk k n_F(k) \int
dp p n_F (p) \left(\frac{1}{(q-i \delta)^2 - 4 k^2}
 + \frac{1}{(q + i\delta)^2 - 4 p^2}\right) \left(\frac{1}{(q-i\delta)^2 - 4 p^2}
  + \frac{1}{(q + i \delta)^2 - 4 p^2}\right) \nonumber \\
&& = 4T \int_0^\infty n^2_F (x) dx = 4 T^2
\left(\log2-\frac{1}{2}\right). \label{a8}
 \eea
\end{widetext}
We emphasize that the integral in the r.h.s. of (\ref{a8}) comes from an
infinitesimally small region where $|k- p| \sim \delta$.

Combining the two contributions to $X_3$, we obtain
\begin{equation}
X_{3}= \left(\log 2 + \frac{1}{2} \right)\frac{2T^{2}}{3}.  \label{a9}
\end{equation}
Collecting (\ref{a4}), (\ref{a7}). and (\ref{a9}), we obtain
\begin{equation}
X=\frac{T^{2}}{3}\left[1 -2 \log {\frac{\Lambda _{\mathrm{f}}}{\Lambda _{%
\mathrm{b}}}}\right].  \label{a10}
\end{equation}
Eq. (\ref{a10}) coincides with Eq. (\ref{40}).

\section{Appendix B}

\label{appB} In this Appendix, we derive the result for $Y=T\sum_{\Omega
}\int dq\Pi _{2k_{F}}^{3}(q,\Omega )$ to logarithmic accuracy. We assume
that $T$ is small, such that $T\ll \Lambda _{\mathrm{b}},\Lambda _{\mathrm{f}%
}$ and that $\Lambda _{\mathrm{b}}\ll \Lambda _{\mathrm{f}}$, and collect
terms logarithmic in $T/\Lambda _{\mathrm{f}}$, and in $\Lambda _{\mathrm{b}%
}/\Lambda _{\mathrm{f}}$. The computational steps are the same as in
Appendix A: we use the fact that $\Pi _{2k_{F}}$ given in (\ref{26}) is the
sum of two terms and split $Y$ into $Y_{1}+Y_{2}+Y_{3}+Y_{4}$, where
\begin{widetext}
\bea
&& Y_1 = T \sum_\Omega \int_0^{\Lambda_{\mathrm{b}}} \frac{dq}{4\pi^3} \log^3 {\frac{\Omega^2 + q^2}{4\Lambda^2_f}}, \label{b11} \\
&& Y_2 = - \frac{6}{\pi^3}  T \sum_\Omega \int_0^{\Lambda_{\mathrm{b}}} dq
\log^2{\frac{\Omega^2 +q^2}{4 \Lambda^2_f}} \int_0^\infty dk
 n_F (k) k  \left(\frac{1}{(q-i\Omega)^2 - 4 k^2} + \frac{1}{(q + i \Omega)^2
  - 4 k^2}\right), \label{b12} \\
&& Y_3 =  + \frac{48}{\pi^3}  T \sum_\Omega \int_0^{\Lambda_{\mathrm{b}}} dq
\log{\frac{\Omega^2 +q^2}{4 \Lambda^2_f}} \left[\int_0^\infty dk
 n_F (k) k  \left(\frac{1}{(q-i\Omega)^2 - 4 k^2} + \frac{1}{(q + i \Omega)^2 - 4 x^2}\right)\right]^2 ,\label{b13} \\
&&Y_4 = -\frac{128}{\pi^3}  T \sum_\Omega \int_0^{\Lambda_{\mathrm{b}}} dq
\left[  \int_0^\infty dx  n_F (x) x \left(\frac{1}{(q-i\Omega)^2 - 4
k^2} + \frac{1}{(q + i \Omega)^2 - 4 k^2}\right)\right]^3.
\label{b14} \eea
\end{widetext}
One can easily make sure last term $Y_{4}$ is non-logarithmic and can be
neglected.

The momentum integral in $Y_{1}$ is infrared divergent. However, we only
need the thermal part of $Y_{1}$. To extract it, we subtract from the
integrand in $Y_{1}$ its value at $\Omega =0$, i.e., $\log ^{3}{\frac{q^{2}}{%
4\Lambda _{\mathrm{f}}^{2}}}$. This makes the momentum integral finite.
Evaluating it and then performing the summation over frequency, we obtain
\begin{equation}
Y_{1}=-\frac{T^{2}}{\pi }\left[ \log ^{2}{\frac{T}{2\Lambda _{\mathrm{f}}}}%
-0.909\log {\frac{T}{\Lambda _{\mathrm{f}}}}+...\right].  \label{b2}
\end{equation}
where dots stand for $O(T^{2})$ terms. The number, 0.909,  as well as other
numbers below are expressed in terms of  convergent 1D integrals.

In $Y_{2}$, the cutoff in the integration over $q_{1}$ is relevant.
Splitting the $q-$integral into $\int_{0}^{\Lambda _{\mathrm{b}}}$ into $%
\int_{0}^{\infty }-\int_{\Lambda _{\mathrm{b}}}^{\infty }$ and evaluating
each of the two terms separately in the same way as in Appendix A, we obtain
\begin{equation}
Y_{2}=\frac{T^{2}}{\pi }\left[ \log ^{2}{\frac{T}{2\Lambda _{\mathrm{f}}}}%
-3.295\log {\frac{T}{\Lambda _{\mathrm{f}}}}+\log ^{2}{\frac{\Lambda _{%
\mathrm{f}}}{\Lambda _{\mathrm{b}}}}-\log {\frac{\Lambda _{\mathrm{f}}}{%
\Lambda _{\mathrm{b}}}}\right].  \label{b3}
\end{equation}
The result from $Y_{3}$ can be readily obtained from the expression for $%
X_{3}$ in Appendix A, as to logarithmic accuracy we can replace $\log {\frac{%
\Omega ^{2}+q^{2}}{4\Lambda _{\mathrm{f}}^{2}}}$ in (\ref{b13}) by $2\log
\left(T/\Lambda _{\mathrm{f}}\right)$. We then obtain
\begin{equation}
Y_{3}=2.386\frac{T^{2}}{\pi }\log {\frac{T}{\Lambda _{\mathrm{f}}}}.
\label{b4}
\end{equation}
Combining (\ref{b2}), (\ref{b3}), and (\ref{b4}), we obtain that all $\log
\left(T/\Lambda _{\mathrm{f}}\right)$ terms are cancelled out, and
\begin{equation}
Y=\frac{T^{2}}{\pi }\left[ \log ^{2}{\frac{\Lambda _{\mathrm{f}}}{\Lambda _{%
\mathrm{b}}}}-\log {\frac{\Lambda _{\mathrm{f}}}{\Lambda _{\mathrm{b}}}}%
\right].  \label{b5}
\end{equation}

Eq. (\ref{b5}) coincides with (\ref{45}).

Another backscattering diagram which could possibly give rise to logarithmic
terms is diagram 3b) in Fig. \ref{fig:third}. For a local interaction, it
reduces to a cube of the Cooper bubble, which in 1D coincides with $\Pi
_{2k_F}$ up to the overall sign. However, one can easily verify that for $%
\Lambda _{\mathrm{b}}\ll \Lambda _{\mathrm{f}}$,
this diagrams does not contain $\log\left(\Lambda _{\mathrm{f}}/\Lambda _{%
\mathrm{b}}\right)$ terms. Indeed, the cutoff induced by the interaction
imposes the restriction on three out of four momenta and frequencies in the
fermionic lines. The 2D integral over the remaining momentum and frequency
involves all six fermionic propagators and is confined to the lower limit.
This implies that all variables are of the same order, and there is no space
for a logarithm.


\begin{thebibliography}{99}
\bibitem{comm_3}  To order $L_{\mathrm{b}}$, we have from  Eqs. (\ref{2})
and (\ref{8b}), ${\tilde g}^2_{1\parallel} + {\tilde g}^2_{1\perp} =
g^2_{1\parallel} + g^2_{1\perp} - 2 ( g^3_{1\parallel} + 3 g^2_{1\perp})
g_{1\parallel} L_{\mathrm{b}}$.
\end{thebibliography}
\end{document}